\documentclass[dvipsnames,twocolumn]{aastex63}

\usepackage{nicefrac}
\usepackage{savesym}
\savesymbol{tablenum}
\usepackage{siunitx}
\usepackage{amsmath}
\usepackage{mathtools}
\usepackage{booktabs}
\usepackage{multirow}
\usepackage{natbib}
\usepackage{dcolumn}
\usepackage{xcolor}
\usepackage{tikz}
\usetikzlibrary{shapes.geometric, arrows}
\usepackage{tablefootnote}
\usepackage{cleveref}
\usepackage{CJKutf8}
\usepackage[utf8]{inputenc}
\usepackage[T1]{fontenc}

\restoresymbol{SIX}{tablenum}
\DeclareSIUnit\parsec{pc}
\DeclareSIUnit\years{yr}
\DeclareSIUnit\Msol{M_{\odot}}
\DeclareSIUnit\Mearth{M_{\earth}}
\DeclareSIUnit\Lsol{L_{\odot}}
\DeclareSIUnit\AU{au}
\DeclareSIUnit\om{\Omega}
\DeclareSIUnit\orb{T_{\mathrm{orb}}}
\DeclareSIUnit\scaleheight{H}
\definecolor{dodgerblue}{rgb}{0.11764706, 0.56470588, 1.}
\definecolor{seagreen}{rgb}{0.18039216, 0.54509804, 0.34117647}
\definecolor{maroon}{rgb}{0.50196078, 0., 0.}

\shortauthors{Pfeil et al.}

\newcommand{\report}[1]{{{#1}}}

\newcommand{\rhog}{\rho_{\mathrm{g}}}
\newcommand{\rhod}{\rho_{\mathrm{d}}}

\newcommand{\amax}{a_{\mathrm{max}}}
\newcommand{\amin}{a_{\mathrm{min}}}

\newcommand{\planck}{\kappa_{\mathrm{P}}}

\newcommand{\athena}{{\normalfont\texttt{Athena++}}}

\newcommand{\radmc}{{\normalfont\texttt{RADMC-3D}}}
\newcommand{\tripod}{{\normalfont\texttt{TriPoD}}}

\begin{document}
\begin{CJK*}{UTF8}{gbsn}



\title{Just a Phase? Weakening Vertical Shear Instability Explains Class II Disk Morphologies:  \\ Simulations with dust coagulation, sedimentation, thermal relaxation, and backreaction}

\author[0000-0002-4171-7302]{Thomas Pfeil}
\affiliation{Center for Computational Astrophysics, Flatiron Institute, 162 Fifth Avenue, New York, NY 10010, USA}
\altaffiliation{The Flatiron Institute is a division of the Simons Foundation.}
\email{tpfeil@flatironinstitute.org}

\author[0000-0003-2966-0419]{Alexandros Ziampras}
\affiliation{University Observatory, Faculty of Physics, Ludwig-Maximilians-Universität München, Scheinerstr. 1, D-81679 Munich, Germany}
\affiliation{Max-Planck-Institut für Astronomie, K{\"o}nigstuhl 17, 69117 Heidelberg, Germany}

\author[0000-0002-8537-9114]{Shangjia Zhang (张尚嘉)}
\affiliation{Department of Astronomy, Columbia University, 538 West 120th Street, Pupin Hall, New York, NY 10027, USA}
\altaffiliation{NASA Hubble Fellowship Program (NHFP) Sagan Fellow}

\author[0000-0001-5032-1396]{Philip J. Armitage}
\affiliation{Center for Computational Astrophysics, Flatiron Institute, 162 Fifth Avenue, New York, NY 10010, USA}
\affiliation{Department of Physics and Astronomy, Stony Brook University, Stony Brook, NY 11794, USA}

\author[0000-0002-2624-3399]{Yan-Fei Jiang (姜燕飞)}
\affiliation{Center for Computational Astrophysics, Flatiron Institute, 162 Fifth Avenue, New York, NY 10010, USA}

\correspondingauthor{Thomas Pfeil}

\begin{abstract}
The Vertical Shear Instability (VSI) is known to create turbulence and strong vertical mixing in protoplanetary disks if thermal relaxation is sufficiently fast.
In simulations where this condition is met, VSI can loft dust particles to large altitudes, creating a vertically extended appearance in mock millimeter-wavelength observations, which is inconsistent with the morphology of the majority of observed class II disks.
We present simulations of protoplanetary disks with VSI that are consistent with the observed thin-disk geometries, while maintaining the commonly observed bowl-shaped morphology in scattered light images at micrometer wavelength.
We show that this outcome arises naturally when the effects of dust coagulation, sedimentation, and dust--gas thermal accommodation are taken into account.
Sedimentation-driven coagulation removes large amounts of dust from the disk atmosphere, in the process slowing down the dust-driven cooling of the gas. At the same time, a dense midplane layer of millimeter-sized grains forms, which exerts aerodynamic drag on the gas.
This results in the termination of the VSI's corrugation mode.
Only weak VSI activity remains in the upper and lower hemispheres.
These processes occur on the typical dust growth timescale and suppress strong VSI-induced turbulence within a few hundred thousand years. VSI could thus generally be restricted to the class I evolutionary stages of protoplanetary disks.


\end{abstract}

\keywords{protoplanetary disks --- dust evolution --- hydrodynamics --- methods: numerical}

\section{Introduction} \label{sec:intro}
Multiwavelength observations of the vertical thickness of the dust layer suggest that most, but not all, protoplanetary disks support low levels of turbulence \citep{Jiang2025,Villenave2025a,Antilen2026}. Independent constraints on turbulence, from gas kinematics \citep{Flaherty2020,Hardiman2026} and from the radial profile of substructures \citep{Ruzza2026}, yield consistent results. Taken at face value, these observational inferences imply that {\em none} of the numerous hydrodynamic and magnetohydrodynamic disk instabilities \citep{Lesur2022} typically lead to significant levels of turbulence, at least at the relatively large radii probed by the data. They also raise an obvious question: if disk turbulence is usually weak, what is physically different about those systems where it is stronger? 

The tension between observations of low turbulence, and theoretical predictions of disk instabilities, is most apparent in the case of the Vertical Shear Instability \citep[VSI;][]{Urpin1998,Nelson2013}, which can generate strong vertical flows that mix dust particles up to several gas scale heights into the disk's atmosphere \citep{Flock2020,Blanco2021, Dullemond2022,Flores-Rivera2025,Flores-Rivera2026}. 
\report{Whether the VSI's coherent, wavelike modes themselves \citep{Svanberg2022,Ogilvie2025}, can be classified as turbulence in the classical sense remains unclear. In this letter, we are mostly interested in their ability to perturb the disk's dust layer. For simplicity, we will nonetheless refer to the related non-Keplerian, non-laminar velocity field as ``turbulence''.}
The simplest model system that realizes the VSI -- a vertically isothermal disk in which cooling is effectively instantaneous -- yields particularly strong vertical motions \citep{Nelson2013,Richard2016,Manger2018,Manger2021,Barraza-Alfaro2024,Barraza-Alfaro2025,Lesur2025}, and can be ruled out. This has led to consideration of physical effects that may alter VSI predictions at leading order. Linear analyses \citep{Lin2015,Malygin2017,Pfeil2019,Fukuhara2021} and simulations \citep{Pfeil2021,Manger2021,Fukuhara2023,Pfeil2023,Fukuhara2024,Pfeil2024} show that the VSI is sensitive to the cooling time, and might only be present if dust particles are small and cooling times are fast. In addition to this thermodynamic effect, the two-way dynamical coupling of dust and gas can also weaken VSI\report{-driven motions} \citep{Schafer2020,Schafer2022,Huang2025a}. These results imply that the simplest VSI model that can fairly be compared to observations is not so simple: it must include at a minimum the combined effects of dust coagulation, sedimentation, backreaction (drag forces transferring momentum from dust to gas, and vice versa), and thermal relaxation. 
The latter, in principle, relies on complex radiation hydrodynamical modeling \citep[RHD,][]{Stoll2016,Zhang2024}. Nevertheless, given the already complex framework outlined above and the expectation that the VSI primarily thrives in optically thin conditions \citep{Lin2015}, we can reasonably well substitute full radiative transfer with an approximate yet accurate cooling model that captures both the radial--vertical thermal structure obtained through RHD as well as the relaxation rate toward said structure.

The \athena{} \citep{Stone2020, Huang2022} hydrodynamics code with the \tripod{} dust coagulation model \citep{Pfeil2024b} includes a nearly self-consistent treatment of the key VSI physics\footnote{The main exception is the linkage between the resolved scales of VSI turbulence, and the effects of that turbulence on vertical mixing and particle collision velocities. This is a difficult multiscale problem that we do not attempt to solve.}. In particular, the use of \tripod{} allows for a self-consistent calculation of the thermal properties of the dust in each grid cell from the evolving local dust size distribution. Here, we use this computational machinery to conduct axisymmetric hydrodynamic simulations of protoplanetary disks, assess the predicted strength of the VSI within them, and compare those results qualitatively against observations. The paper is structured as follows. We introduce the methods (\tripod{}, \athena{}) in \autoref{sec:methods}. In \autoref{sec:results}, we present our simulations and results. To demonstrate the influence of the individual physical effects, we run a series of simulations with increasing complexity; starting from the simplest possible isothermal setup, to simulations with dust-gas cooling, and finally the full range of physics included in our code. We conclude and discuss our findings and their implications in \autoref{sec:conclusions}.

\section{New Methods} \label{sec:methods}
We conduct axisymmetric hydrodynamic simulations of protoplanetary disks with the \athena{} multifluid module \citep{Stone2020, Huang2022}. Dust coagulation is included using the \tripod{} model, which evolves a power-law dust size distribution in every grid cell of our simulations \citep{Pfeil2024b}\footnote{Originally designed for vertically integrated simulations of protoplanetary disks, \tripod{} has been adapted and tested for radial-vertical setups (i.e., axisymmetric or three-dimensional simulations), in \cite{Eriksson2026}.}. In this prescription, the entire polydisperse size distribution is described by three evolving parameters $\amax$ (the maximum particle size), $\rho_\mathrm{d,0}$ (the density of particles of sizes within $[\amin, \sqrt{\amin \amax}]$), and $\rho_\mathrm{d,1}$ (the density of particles of sizes within $[\sqrt{\amin \amax}, \amax]$). Together, these define the power-law exponent of the dust size distribution $\rho_\mathrm{d}(a)\propto a^{q+3}$, as
\begin{equation}
    q = 2\frac{\report{\ln}\left(\rho_\mathrm{d,1}/\rho_\mathrm{d,0}\right)}{\report{\ln}\left(\amax /\amin\right)} - 4\ ,
\end{equation}
such that the full distribution can be reconstructed via
\begin{align}
   \rho_\mathrm{d}(a)= \begin{cases}
    \dfrac{\rho_{\text{d,tot}}(q+4)}{\amax^{q+4} - \amin^{q+4}} a^{q+3} & \text{for } q\neq -4 \\[15pt]
    \dfrac{\rho_{\text{d,tot}}}{\report{\ln}(\amax)-\report{\ln}(\amin)}\dfrac{1}{a} & \text{for } q=-4\ .
  \end{cases}
\end{align}
We implement the thermal coupling of dust and gas by calculating the local thermal relaxation times of the gas, given the local dust size distribution. Following \cite{Barranco2018},
\begin{align}
    t_\mathrm{thin}^\mathrm{NLTE} &= 2t_{||}\left[1-\sqrt{1-\frac{4t^2_{||}}{t_\mathrm{g}^\mathrm{coll} t_\mathrm{d}^\mathrm{rad}}}\right]^{-1} , \\ 
    \frac{1}{t_{||}} &= \frac{1}{ t_\mathrm{d}^\mathrm{rad}}+\frac{1}{t_\mathrm{d}^\mathrm{coll}}+\frac{1}{t_\mathrm{g}^\mathrm{coll}} \nonumber \ .
\label{eq:tNLTE}
\end{align}
with the dust radiative timescale given by
\begin{equation}
    t_\mathrm{d}^\mathrm{rad} \approx \frac{C_\mathrm{d} }{\displaystyle 16 \,\sigma_\mathrm{SB}\, T^3 } \left(\frac{\int \rhod(a)\planck(\amax,q,T) \, \mathrm{d}a}{\int \rhod(a) \, \mathrm{d}a}\right)^{-1},
\end{equation}
where $C_\mathrm{d}=\SI{8e6}{\mathrm{erg} \per \gram \per \kelvin}$ is the specific heat at constant volume for the dust grains, $\sigma_\mathrm{SB}$ is the Stefan-Boltzmann constant, $T$ is the dust temperature, and $\rhod(a)$ is the dust size distribution. This timescale is sensitive to the local dust size distribution through the dust-density averaged Planck mean opacities. 
We use a three-dimensional dust opacity table on which we perform a trilinear interpolation in every grid cell given the local dust size distribution $\amax,\, q$ and temperature $T$ (see \autoref{app:OpTab}).
Note that this timescale is orders of magnitude shorter than the other relevant timescales at any given time for the performed simulations. Our results are thus not sensitive to the employed values of the dust opacity or $C_\mathrm{d}$, since cooling is limited by the much slower energy transfer between dust and gas.

The gas-dust thermal accommodation timescales for the dust ($t_\mathrm{d}^\mathrm{coll}$) and gas ($t_\mathrm{g}^\mathrm{coll}$) are given by 
\begin{align}
    t_\mathrm{g}^\mathrm{coll} &= \frac{\gamma}{\gamma-1}\frac{1}{n_\mathrm{S}\sigma_\mathrm{S}\Bar{v}_\mathrm{g}},  
    \label{eq:tgcoll}\\
    t_\mathrm{d}^\mathrm{coll} &= \left(\frac{\rhod}{\rhog}\right)\left(\frac{C_\mathrm{d}}{\mathrm{C_P}}\right) t_\mathrm{g}^\mathrm{coll}\ ,
    \label{eq:tdcoll}
\end{align}
where $C_\mathrm{P}$ is the gas' specific heat at constant pressure and $\gamma=1.43$ is the gas' heat capacity ratio. We define the representative number density $n_\mathrm{S} =\rhod/\left(\nicefrac{4}{3}\, \pi \rho_\mathrm{m} a_\mathrm{S}^3\right)$ and the representative collisional cross-section $\sigma_\mathrm{s}=\pi a_\mathrm{S}^2$, with $\rho_\mathrm{m}=\SI{1.67}{\gram \per \cubic \centi \meter}$ being the interior density of the dust grains, and $\bar{v}_\mathrm{g}$ is the mean gas molecular velocity.
The Sauter-mean particle size \citep{Sauter1926} for the \tripod{} size distributions in our hydrodynamic simulations is given by
\begin{align}
a_\mathrm{S} =
\begin{cases}\vspace{2mm}
    \dfrac{\amax-\amin}{\report{\ln}(\amax)-\report{\ln}(\amin)} &\text{if $q=-3$} \\ \vspace{2mm}
    \dfrac{\amax\amin}{\amax-\amin} \report{\ln}\left(\dfrac{\amax}{\amin}\right)&\text{if $q=-4$} \\  
    \left(\dfrac{q+3}{q+4}\right) \dfrac{\amax^{q+4}-\amin^{q+4}}{\amax^{q+3}-\amin^{q+3}} &\text{otherwise,}
\end{cases}
\label{eq:Sauter}
\end{align}
in every grid cell.
Note that the above definitions \ref{eq:tgcoll}, \ref{eq:tdcoll}, and \ref{eq:Sauter} of the collisional accommodation timescale are exact for a power-law size distribution. 

Given this framework, we relax temperature perturbations $T^{(n)}-T_\mathrm{eq}$ toward the equilibrium temperature profile $T_\mathrm{eq}$ in our simulations via
\begin{equation}
    T^{(n+1)} = T_\mathrm{eq} + (T^{(n)}-T_\mathrm{eq})\exp\left(-\frac{\Delta t}{t_\mathrm{thin}^{\text{NLTE}}}\right);\label{eq:relax}
\end{equation}
where $\Delta t$ is the $(n)^\mathrm{th}$ simulation timestep.
We explore the effects of two different equilibrium temperature structures: vertically isothermal ($T_\mathrm{eq}(R)$) and vertically stratified ($T_\mathrm{eq}(R,z)$. 
The radial power-law temperature profile has an exponent of $\beta_T=-0.5$ (see \autoref{app:TStruc} for details).
We prescribe a vertically hydrostatic \cite{LyndenBell1974} density profile (see \autoref{app:DenStruc}) with power-law exponent $\beta_\Sigma=-1$ and a disk mass of \SI{0.05}{M_\odot}. 

We conduct axisymmetric simulations using the Harten–Lax–van Leer–Contact Riemann solver with RK2 time stepping and linear reconstruction on a spherical-polar grid. The resolution is $\SI{64}{cells \per scale height}$ (at \SI{50}{\AU}) on a domain from \SIrange{25}{100}{\AU} in radius and $\pm 5$ pressure scale heights ($H/R(\SI{50}{\AU})\approx 0.07697$) in the polar direction. This results in a $3120(r)\times 637(\vartheta)$ grid. The initial dust-to-gas density ratio is constant throughout the disk with $\varepsilon=0.01$. 
The dust initially follows an MRN size distribution \citep*{Mathis1977} with a minimum size of \SI{0.1}{\micro\meter} and a maximum size of \SI{1}{\micro\meter}.

\begin{figure*}[ht]
    \centering
    \includegraphics[width=\linewidth]{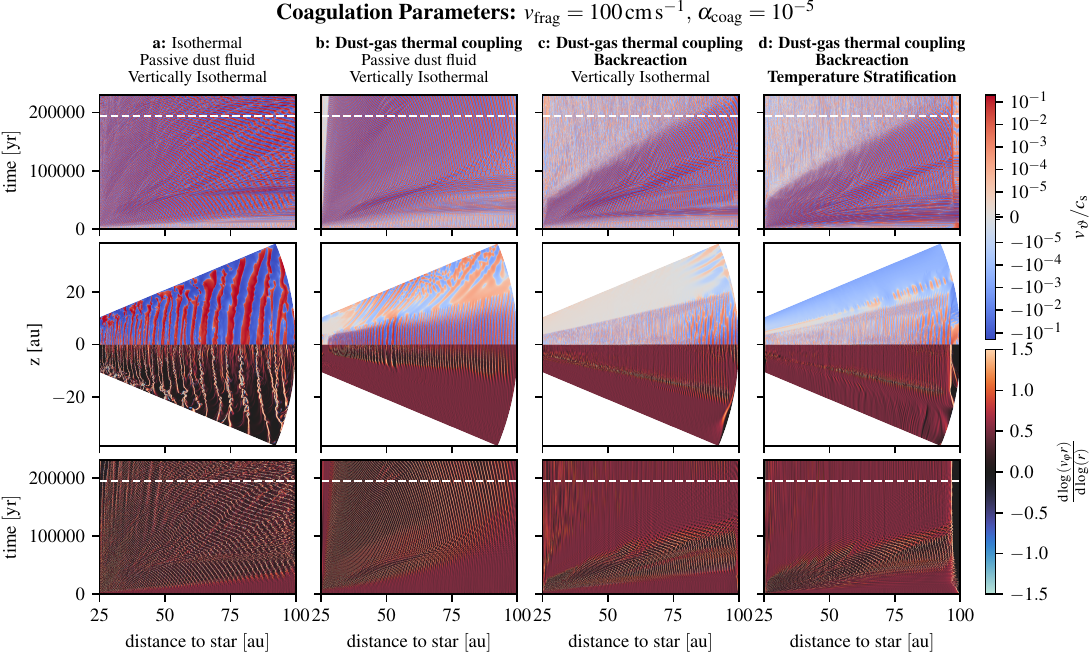}
    \caption{Space-time diagrams visualizing the effects of different physical processes on the time evolution of the Vertical Shear Instability. The top row shows the evolution of the midplane polar Mach number in the disk's midplane, which indicates the presence of the VSI's corrugation mode. The third row shows the radial gradient of specific angular momentum at one pressure scale height distance from the midplane. A flat profile (black) indicates the saturated VSI modes.
    The central row shows a simulation snapshot toward the end of the simulation (dashed horizontal lines in row one and three), showing the respective quantities throughout the simulation domain.}
    \label{fig:vfr100_gas}
\end{figure*}

\begin{figure*}[ht]
    \centering
    \includegraphics[width=\linewidth]{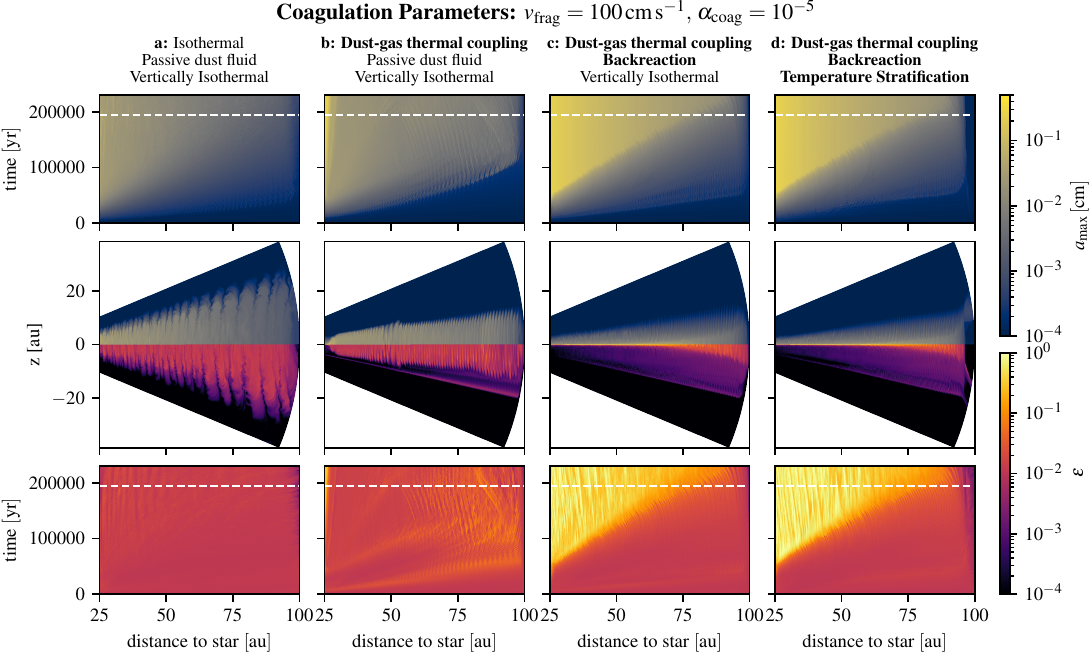}
    \caption{Space-time diagrams visualizing the effects of different physical processes on the time evolution of the Vertical Shear Instability and its effect on the dust particles. The top row shows the midplane maximum particle size. 
    The bottom row shows the evolution of the midplane dust-to-gas ratio.
    The central row shows a simulation snapshot toward the end of the simulation (dashed horizontal lines in row one and three), showing the respective quantities throughout the simulation domain.}
    \label{fig:vfr100_dust}
\end{figure*}

\section{Results} \label{sec:results}
\begin{figure}[ht]
    \centering
    \includegraphics[width=\linewidth]{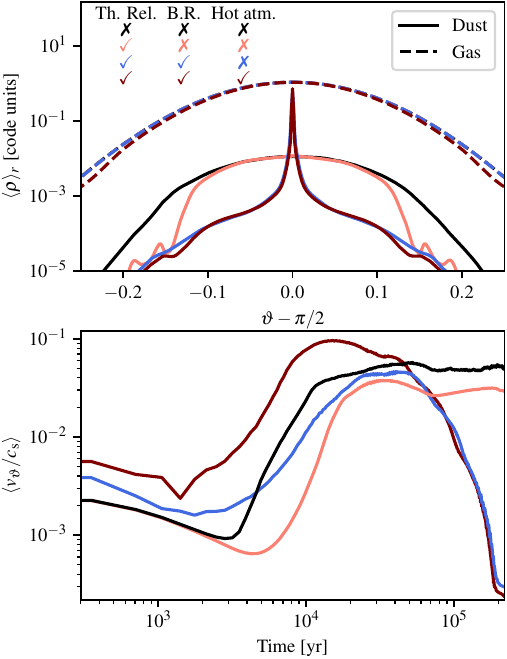}
    \caption{Radially averaged vertical density profiles in our four simulations (dashed lines show gas densities, solid lines dust densities). While the ideal VSI simulation and the simulation with thermal relaxation but no backreaction maintain a vertically extended dust layer (black and light red), we find strongly sedimented midplane dust layers in the simulations that consider backreaction (blue and dark red). The legend shows the included physical effects in the respective simulation (thermal relaxation, backreaction, hot atmosphere)}
    \label{fig:densities}
\end{figure}
\subsection{toward a Self-Consistent Simulation}
To illustrate the effects of the individual physical effects of dust-gas cooling, backreaction, and vertical temperature stratification, we present four simulations with increasingly comprehensive physics. 
The simulations employ a fragmentation velocity of \SI{100}{\centi\meter\per\second} and assume that turbulent particle collisions are mostly negligible, with a turbulent $\alpha_\mathrm{coag}=10^{-5}$. This means grain-grain collisions are initially driven by Brownian motion (in the midplane, where the initial Stokes numbers are very small), relative drift, and relative sedimentation \citep[in the atmosphere, where the initial Stokes numbers are already larger for micron-sized particles, see][]{Zsom2011,Eriksson2026}.

\paragraph{Ideal VSI}
We designate the simulation conducted with an isothermal equation of state, without dust backreaction, and in a vertically isothermal setup, as ``ideal VSI''. This setup is essentially identical to that of \cite{Nelson2013} (except for the exponentially tapered disk, see \autoref{app:DenStruc}). In this simplest possible global VSI model, persistent corrugation motions develop across the entire simulation domain, as shown in column (a) of  \autoref{fig:vfr100_gas}. 
The turbulence saturates everywhere in the domain after $\sim \SI{2e5}{\years}$ when bands of constant angular momentum have formed, showing up as dark/black bands in the lower half of column (a) of \autoref{fig:vfr100_gas}. 
These bands also show up as black stripes in the space-time plot in the lower half of the figure, which shows the angular momentum gradient at one pressure scale height.
At this point, emerging vortices directly tap into the baroclinic term and start to grow. \cite{Klahr2026} described this effect and related it to the sea-breeze effect in geophysical fluid dynamics \citep{Holton2012}. It results in the formation of giant vortices in the $R-z$ plane. As it is usually observed only in axisymmetric simulations we terminate the simulation at this point.

The strong vertical motions loft even large dust particles into the upper layers of the disk, meaning the dust-to-gas density ratio in the disk's midplane remains relatively constant throughout the simulation, despite the dust coagulation process increasing the particles' size over time, as shown in column (a) of \autoref{fig:vfr100_dust}. 
Coagulation proceeds in the typical sedimentation-driven pattern that is present if turbulence is weak. Particle collisions are initially dominated by relative settling in the upper disk atmosphere, where the growing grains sweep up the smaller particles on their way to the midplane. The growth therefore manifests as a rain-out of large, coagulating particles from the upper atmosphere of the disk. Finally, the largest grains reach the fragmentation barrier driven by the weak turbulence in the midplane. 
The maximum grain size reaches a steady state after $\sim \SI{e5}{\years}$ throughout the disk with $\amax\sim \SIrange{0.1}{2}{\milli \meter}$ in the midplane.

\paragraph{Effect of Finite Cooling Times}
Column (b) of \autoref{fig:vfr100_gas} shows a simulation in which the thermal relaxation timescale of the gas is calculated based on the locally evolving dust size distributions. The dust fluids themselves are, however, still treated as passive, i.e., no drag force is exerted on the gas. This setup is similar to the simulations in \cite{Fukuhara2025}, and \cite{Pfeil2021, Pfeil2023, Pfeil2024}, with the important difference that the grain size distribution is evolving over time as a consequence of coagulation and fragmentation in every grid cell.
This simulation reaches an equilibrium state, where coagulation and sedimentation have depleted the upper atmosphere of the disk, leading to long thermal relaxation times (\autoref{fig:vfr100_dust}, column (b)). Hence, VSI activity has completely ceased in the upper layers, similar to the earlier studies. The VSI turbulence is thus confined to a small area around the disk midplane, where it drives strong enough vertical mixing to maintain an extended dust layer.
Midplane dust-to-gas ratios and particle sizes show only minor deviations from the fully isothermal simulation due to this efficient vertical dust transport. 

As pointed out by \cite{Fukuhara2024}, this situation is only sustainable as long as the particles have relatively low Stokes numbers. Larger fragmentation velocities can lead to a situation with quenched VSI (see \autoref{app:vfrag}).

\paragraph{Combined Effect of Finite Cooling Times and Drag Forces}
The situation changes drastically if the gas experiences the drag forces exerted by the dust fluids. Column (c) of \autoref{fig:vfr100_gas} and \autoref{fig:vfr100_dust} shows the first radial-vertical simulation of a protoplanetary disk to include the combined effects of dust-gas thermal coupling, coagulation, fragmentation, and drag forces. 
As can be seen in the space-time plots in these figures, VSI activity initially emerges everywhere in the disk, similar to the simplified simulations in columns (a) and (b). 
The initial activity is, however, slightly more vigorous, with stronger initial modes, visible as slightly higher initial velocities (see also \autoref{fig:densities}). The reason for this is the sedimenting dust layer, which provides a strong initial perturbation to the gas, which then undergoes VSI. 
The initially strong VSI activity is however damped as soon as dust coagulation and sedimentation proceed.
The dust density enhancement in the disk midplane due to the ongoing sedimentation-driven coagulation leads to an increase of adverse drag forces that act against the VSI's corrugation modes. This effect has been discussed thermodynamically by \cite{Lin2017} and has also been observed by \cite{Lin2019,Schafer2020, Schafer2022}, although without coagulation or realistic cooling times. These authors found that, despite the termination of the corrugation mode in the disk midplane, VSI could still persist in the form of strong breathing modes in the upper and lower hemispheres of the disk in their isothermal simulations. 
The situation is, however, dramatically different if realistic cooling times are considered. 
As soon as the corrugation modes are terminated in the disk midplane by the dust backreaction, only weak breathing modes remain due to the reduced thermal relaxation in the dust-depleted upper layers. 
Sedimentation-driven coagulation easily overcomes the remaining vertical gas motions and leads to the formation of a highly concentrated dust layer in the disk's midplane. As a consequence, the disk atmosphere becomes highly dust-depleted, with only a low density of micron-sized grains remaining suspended in the gas.
As the dust growth front propagates outwards, VSI is sequentially shut off from the inside out within $\sim \SI{2e5}{\years}$. 
As can be seen in the second panel of column (c) in \autoref{fig:vfr100_dust}, dust-to-gas density ratios reach values of unity and higher, once the initial VSI activity has subsided. We cannot resolve the dynamics of these dust-dominated regions in our global simulations, and hence cannot observe the occurrence of the streaming instability \citep{Youdin2005} or dust clumping. \cite{Schafer2020, Schafer2022} have, however, shown that drag-induced instabilities will eventually take over in the midplane.
A very weak VSI motion persists in the atmosphere of the disk and does not cross over the midplane. It is not able to lift the larger particles into the atmosphere.

\paragraph{Effects of Vertical Temperature Stratification} The temperature structure of protoplanetary disks is mostly determined by the incident stellar irradiation. Small dust particles in the disk atmosphere absorb and scatter most of the incoming energy, which is then transported toward the optically thick midplane. Radiative transfer models usually show super-heated upper disk layers, at 2-3 times the midplane temperature, with a sharp transition around the disk's photosphere.
Works by \cite{Zhang2024} and \cite{Yun2025, Yun2025a} have shown that the additional vertical shear arising from this stratification can drive a very strong VSI originating in the transition layer. Although the resulting modes originate from high up in the atmosphere, they can potentially drive significant perturbations in the disk midplane and lift dust particles up. To investigate this effect in combination with dust evolution, we conduct simulations with a vertical temperature structure that is informed by radiative transfer models using flux-limited diffusion (see \autoref{app:TStruc}).
We note that the simulation presented here represents only one possible vertical temperature structure, which enters via the equilibrium temperature in the thermal relaxation step.
The atmospheric temperature of this model is $\sim 3$ times higher than the midplane temperature, which is the same as the temperature in our vertically isothermal models. The temperature transition occurs at $\sim 4.6$ pressure scale heights.
In this model, dust sediments out of the transition layer before significant VSI growth can occur. The outcome is therefore almost identical to the vertically isothermal simulation with the same assumed dust physics (see column (d) in \autoref{fig:vfr100_gas} and \autoref{fig:vfr100_dust}). 
Temperature stratification is therefore not always able to keep the VSI active. 
Different temperature structures, however, can create very strong VSI turbulence, which is inconsistent with the often observed razor-thin disk morphology (see \autoref{app:TstrucSims})---especially with the transition is wider and the profile more bowl-shaped.

\paragraph{Averaged disk quantities} \autoref{fig:densities} shows the radially averaged dust density at the end of our simulations and the time evolution of the averaged polar Mach numbers.
Ideal VSI and the simulation without backreaction can be seen to maintain a very thick dust layer. In simulations with backreaction, we find a highly concentrated midplane dust layer where dust-to-gas ratios approach unity on average. The upper layers in these simulations are severely depleted by a factor of $\sim 100$.
In the lower panel of \autoref{fig:densities}, we can see that the simulations with backreaction begin to develop VSI slightly faster due to the initial perturbation caused by the sedimenting dust layer. Turbulent velocities are, however, sharply declining in these models after just \SI{5e4}{\years}, while the simulations without backreaction maintain the VSI turbulence.

\begin{figure*}[ht]
    \centering
    \includegraphics[width=\linewidth]{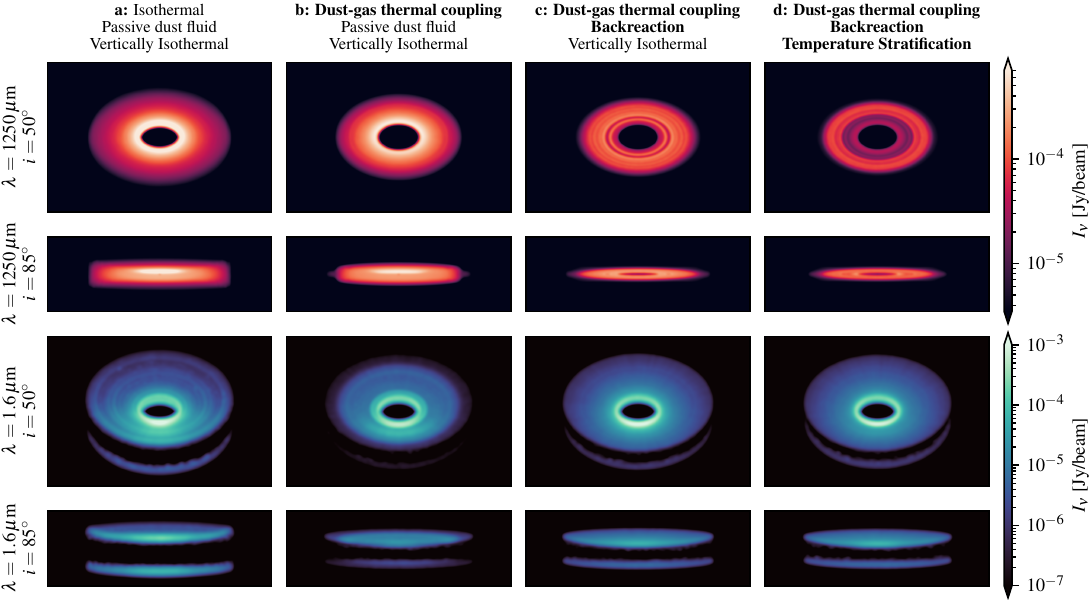}
    \caption{\radmc{} models of our simulation data at \SI{1.25}{\milli \meter} dust continuum (red color maps, row one and two), and \SI{1.6}{\micro \meter} wavelength, showing scattered light (blue color maps, rows three and four). Dust continuum images have been convolved with a beam identical to the observations published in \cite{Villenave2022}; scattered light images have been convolved with a circular Gaussian beam with a \SI{30}{mas} FWHM. The innermost \SI{1}{\AU} of the images has been masked before the beam convolution.}
    \label{fig:RADMC}
\end{figure*}
\subsection{Synthetic Observations}
To connect our simulation results to scattered light and dust continuum observations of protoplanetary disks, we create mock observations with \radmc{} \citep{Dullemond2012}. We first reconstruct a dust size distribution with 16 bins based on the \tripod{} dust parameters in every given grid cell. To make the calculations tractable, we down-sample our simulation domain and post-process only every fourth radial and polar grid cell. For the azimuthal domain, we define a regular grid with 128 cells from $0$ to $2\pi$, resulting in a density model with $780(r)\times160(\vartheta)\times 128(\varphi)\times 16(a)$ cells.
We calculate absorption and scattering opacities using the \texttt{dsharp\_opac} Python module for the default particle properties used in DSHARP \citep{Birnstiel2018}. 
After conducting thermal Monte-Carlo simulations to determine the disks' temperatures using $10^7$ photon packages, we calculate images at inclinations $i=\SIlist{50;85}{\degree}$, each at wavelength $\lambda=\SIlist{1.6;1250}{\micro \meter}$. 
To make the results more comparable to actual observations, we convolve the models at \SI{1.25}{\milli\meter} with a beam identical to the observations by \cite{Villenave2022}; and the scattered light images with a circular Gaussian beam with \SI{30}{mas} FWHM after masking the central star; both at a distance of \SI{147}{\parsec}.
The results are shown in \autoref{fig:RADMC}.

\subsubsection{Dust Continuum}
\begin{figure}
    \centering
    \includegraphics[width=\linewidth]{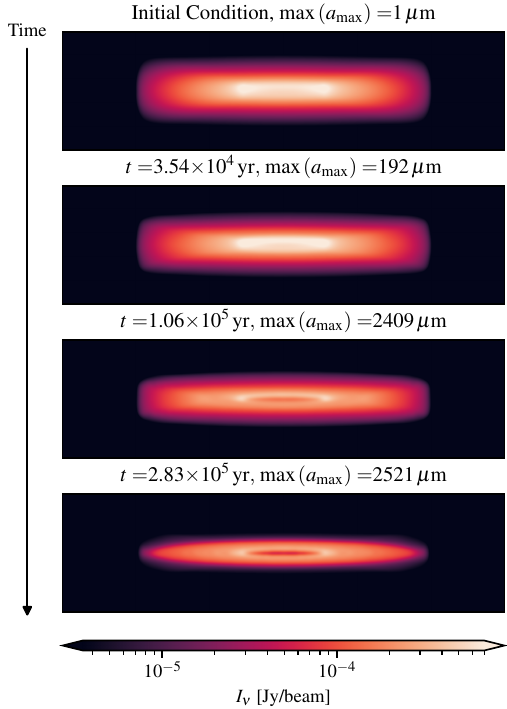}
    \caption{Disk evolution timeseries seen at $\lambda=\SI{1.25}{\milli\meter}$ dust emission, convolved with the same beam as \autoref{fig:RADMC}. Dust is initiated with a constant dust-to-gas ratio. Over time VSI turbulence shuts down, which results in a razor-thin disk of large grains.}
    \label{fig:RADMC_Timeseries}
\end{figure}
Ideal, isothermal VSI results in a vertically extended disk structure as can be seen in column (a) of \autoref{fig:RADMC}. 
The large amount of dust in the atmosphere also results in the enhanced absorption of stellar irradiation in the upper layers, which increases the dust continuum emission compared to a thinner disk. 
As discussed by \cite{Dullemond2022}, this disk morphology is inconsistent with the majority of Atacama
Large Millimeter/submillimeter Array (ALMA) observations of class II disks, which predominantly show razor-thin geometry \citep{Villenave2022, Villenave2025a}.
Including the effects of dust-gas cooling and varying dust opacities reduces the turbulence strength, as discussed above, but still allows for a moderately turbulent midplane. The resulting morphology is still vertically extended, as can be seen in column (b) of \autoref{fig:RADMC}.
Including the dust feedback renders the disk's midplane mostly laminar at the end of the simulation, which leads  to the aforementioned dust-dominated midplane layer and a weakly turbulent atmosphere. The dust continuum emission, shown in column (c) of \autoref{fig:RADMC}, appears razor-thin, consistent with observed disks such as Oph 163131 \citep{Villenave2022}.
Including the physics of dust coagulation, sedimentation, variable dust opacities, and diminishing dust-gas collisional coupling thus naturally results in razor-thin dust disks over the course of a few \SI{100}{\kilo\years}. This resolves the tension between observations and the unrealistic ideal VSI.


The simulation with vertical temperature stratification results in a similarly thin midplane dust layer and hence leads to a similarly thin appearance in the dust continuum emission. This simulation appears slightly fainter in the dust continuum due to the slightly lower total dust mass, which results from the different vertical density stratification (we used the same midplane density across all models).

\subsubsection{Scattered Light}
Scattered light images reveal the presence of micron-sized grains in the upper atmosphere of protoplanetary disks. 
Sedimentation-driven coagulation, as present in our simulations, removes large amounts of small dust from the upper atmosphere of the disk. As can be seen in the upper panel of \autoref{fig:densities}, dust-to-gas density ratios above approximately two pressure scale heights are reduced by about a factor of 100 and further decline with distance to the midplane due to this effect. As discussed above, the resulting collisional decoupling of dust and gas results in the termination of the VSI's corrugation motions and results in the razor-thin millimeter-wavelength observations presented in the previous section.
To check whether our models with severely dust-depleted atmospheres are still consistent with scattered light images of protoplanetary disks, we have created \radmc{} models at \SI{1.6}{\micro\meter} wavelength, shown in the lower two rows of \autoref{fig:RADMC}.

In column (a) of \autoref{fig:RADMC}, we present images made from our ideal VSI simulation. As the entire simulation domain is turbulent, we can also see signs of VSI in the disk surface because of the corrugated layer of micron-sized grains in the upper atmosphere.
Small grains are continuously mixed up and resupplied by fragmentation in the disk's midplane. 
The presence of larger grains in the upper disk layers results in a prominent increase of brightness on the observed side of the disk. This can be attributed to forward scattering by large dust particles.

In column (b), we show the simulation including the thermal coupling of dust and gas. As turbulence is confined to a region around the disk's midplane, we also find a slightly more compact appearance in the edge-on scattered light image. The prevalent ring structures seen in the vertically isothermal model have mostly disappeared in the image due to the weaker VSI action. 
The effect of strong forward scattering is present but slightly subdued due to the reduced presence of large grains. 

In the simulations including backreaction in columns (c) and (d), we still find the characteristic vertically extended morphology in the scattered light images, despite the strongly depleted upper atmosphere.

\subsection{An Evolutionary Sequence of VSI Turbulence}
We illustrate the evolutionary pattern of the disk in synthetic dust continuum images of the different time steps of simulation (c) in \autoref{fig:RADMC_Timeseries}. As can be seen, the evolution begins with a vertically extended dust layer of micron-sized grains. The vertical extent of the structure is initially maintained by strong VSI turbulence but becomes gradually weaker as the dust grows and sediments. After $\sim \SI{100}{\kilo\years}$, VSI in the inner disk has also already subsided, while the outer regions are still moderately turbulent. At $\sim \SI{280}{\kilo\years}$, turbulence has been quenched throughout the entire disk midplane, resulting in the razor-thin geometry, shown in \autoref{fig:RADMC}, column (c).

\section{Discussion}
The class I stage of Young Stellar Object evolution is likely to be associated with an initial phase of strong turbulence, driven by gravitational instability \citep{Kratter2016} and/or infall. Following this phase, our results suggest that the resulting small dust particles can give rise to a short-lived phase of VSI-induced turbulence that lasts only a few hundred thousand years until sedimentation-driven coagulation suppresses the VSI in the upper disk atmosphere. The timescale for the persistence of strong VSI turbulence in the disk's midplane is approximately one local dust coagulation/sedimentation timescale. Observationally, our predicted evolution in the strength of the VSI -- driven by dust coagulation and feedback -- implies that razor-thin disk morphologies arise naturally in the early stages of class II observations that probe more evolved disk populations. Therefore, observational works such as those of \cite{Villenave2025a}, may for the most part have missed an earlier VSI-active era.

\subsection{Vertical Temperature Stratification}
\label{sec:dissT}
Vertical temperature stratification can enhance VSI activity under certain circumstances. One vertical temperature structure, that leads to a result consistent with observations, has been presented in the main part of this work. 
We caution, however, that this particular setup only represents one possible realization of a disk structure. The appendix shows the results from two additional simulations, where the transition between cool midplane and hot upper layers is moved to lower height. Previous results show that this configuration is more favorable for strong VSI. We find that if the transition occurs closer to three pressure scale heights, a very strong VSI leads to the formation of bands of constant specific angular momentum and a vertically extended layer of millimeter-sized grains. However, even these initially highly turbulent models become mostly quiescent after a few hundred \si{\kilo\years}. We therefore conclude that even with a favorable temperature structure, VSI is likely a  temporary phase in protoplanetary disk evolution, though the exact duration has some sensitivity to the vertical structure.

Going beyond models with a fixed vertical temperature structure, the changing dust scale height would be expected to influence the vertical temperature structure of the disk over time. As the dust sediments, the temperature transition region would move toward the midplane. We show the post-processed vertical temperature structure at the first and last time stamps in \autoref{fig:RADMC_Temp} of the appendix. 
We do not account for this in our simulations.

These effects must be studied in more detail in the future and require self-consistently calculated temperature structures using radiation-hydrodynamic simulations. We will investigate this in future work, utilizing the dynamically evolving opacities developed for this study in combination with the discrete-ordinates radiative transfer scheme in \athena{} \citep{Jiang2021,Baronett2026}.

\subsection{Effects of the Fragmentation Velocity}
The grain fragmentation velocity we chose here, \SI{100}{\centi \meter \per \second}, is at the lower end of most estimates for the typical break-up velocity of dust particles in protoplanetary disks based on laboratory experiments \citep{Blum2018} and molecular dynamics simulations \citep{Morrissey2026}. 
Since collisional decoupling in the atmosphere due to efficient grain growth and settling is the main reason for the declining VSI activity in our simulations, applying larger values of $v_\mathrm{frag}$ would lead to even weaker VSI on shorter timescales. 
Our simulations therefore represent a favorable scenario for VSI.
We present two simulations with a fragmentation velocity of \SI{400}{\centi \meter \per \second} in \autoref{app:vfrag}.

\subsection{Three-dimensional Effects: Flow Structure Formation}
Although VSI in our simulations subsides after a short time, it might still contribute to vortex formation in the disk's midplane. Simulations by \cite{Manger2018} \report{and} \cite{Pfeil2021} have shown vortex formation to set in within $\sim 100$ orbits, after which the resulting small structures merge and form vertically extended, long-lived vortices. 
It has not been investigated to date how varying cooling times influence this process. Subcritical Baroclinic Instability has been discussed as a mechanism sustaining and enhancing vortices at cooling times comparable to the orbital timescale \citep{Klahr2003,Petersen2007b}. 
As dust coagulation progresses and cooling times become longer over time due to the depletion of small grains, protoplanetary disks can evolve into this cooling time regime with potential consequences for the formation and longevity of VSI-induced vortices. We will investigate this process in future work.

\report{VSI turbulence in two-dimensional simulations furthermore behaves differently from three-dimensional setups. Forming bands of constant angular momentum and the related sea-breeze effect are enhanced in axisymmetric simulations. Non-axisymmetric perturbations are non-existent by construction in these models, which likely results in different power spectra and energy transport throughout the simulation.}

\subsubsection{Comparison to Previous Work}
Previous works on the evolution and appearance of VSI active disks have investigated the effects of different physical effects.
\cite{Pfeil2021} discussed the effects of dust-gas thermal decoupling in the atmosphere through gas-only simulations, assuming a well-mixed micron-sized dust background. They showed that VSI turbulence is capped in the upper layers due to long cooling times (our simulation b). 
\cite{Fukuhara2021,Fukuhara2023,Fukuhara2024,Fukuhara2025} further investigated the effects of the collisional cooling time for various Stokes numbers and devised a semi-analytic model to derive the VSI-active zones in protoplanetary disks, followed by hydrodynamic simulations with fixed-size dust fluids.
Work by \cite{Lin2017} and by \cite{Lin2019} discussed the effects of drag forces on the VSI and concluded that strong feedback can inhibit turbulence in the disk midplane. \cite{Schafer2020,Schafer2022} confirmed this in isothermal simulations with static mesh refinement and found VSI activity can still be maintained in both hemispheres, separated by a dust-dominated midplane layer.
\cite{Dullemond2022} demonstrated through \radmc{} post-processing of ideal VSI simulations that strongly corrugated dust layers caused by ideal VSI are not consistent with ALMA observations. They also discussed the effects of a dust-depleted atmosphere and showed that even a depletion factor of 100 can maintain the typical bowl-shaped appearance in scattered light images. Our new simulations with dust coagulation now show that such a structure is maintained even if sedimentation-driven coagulation removed the large dust from the atmosphere.

In \cite{Pfeil2023}, we investigated the effect of dust coagulation in a highly simplified manner by using a fixed cooling time distribution calculated from \texttt{DustPy} \citep{Stammler2022} models. These studies could therefore not account for any dust dynamics or ongoing coagulation. 
To improve on this, we added dust as passive scalars to our simulations in \cite{Pfeil2024} to study the effect of dust sedimentation. Sedimentation-driven coagulation and backreaction, which we show here are of great importance, were not included in these simulations.

Our new work solves these discussed issues in a single, physically consistent simulation. Sedimentation-driven coagulation, collisional decoupling in the upper atmosphere, and dust-dominated midplane layers arise naturally and self-consistently in our simulations.
Including a more realistic, self-consistent vertical temperature structure in our simulations will be the next step.

\section{Conclusions} \label{sec:conclusions}
Dust coagulation is essential for studies of instabilities that depend on the thermal relaxation timescale of the gas and on dust-gas dynamic feedback. Including these effects in our simulations, we have shown that the VSI is likely a short-lived phenomenon in the evolution of protoplanetary disks that likely terminates before the disks enter their class II stage. 
While the depletion of dust in the upper atmosphere leads to the termination of VSI activity high up in the disk, drag forces terminate the VSI's corrugation mode in the midplane once the Stokes numbers and the dust-to-gas ratios have increased due to sedimentation-driven coagulation. 

Applied to observations, our results suggest that the typical morphology of disks, that is seen in dust continuum and scattered light observations, may be a consequence of VSI physics. \report{The weakening of VSI allows the dust sedimentation to proceed unhindered and} explains the geometrically thin appearance of most class II protoplanetary disks in millimeter-wavelength dust continuum observations; \report{a} low-density halo of micron-sized grains that remains suspended in the disk's atmosphere creates the typical bowl-shaped appearance of scattered light observations.

\section*{Acknowledgments}
AZ acknowledges funding from the European Union under the European Union's Horizon Europe Research and Innovation Programme 101124282 (EARLYBIRD). Views and opinions expressed are those of the authors only.
Support for this work was provided by NASA through the NASA Hubble Fellowship grant
\#HST-HF2-51568 awarded by the Space Telescope Science Institute, which is operated by the
Association of Universities for Research in Astronomy, Inc., for NASA, under contract
NAS5-26555.

\section*{Software}
\begin{itemize}
    \item \athena{} \citep{Stone2020}  multifluid module \citep{Huang2022}
    \item \tripod{} \citep{Pfeil2024b}
    \item \radmc{} \citep{Dullemond2012}
    \item \texttt{dsharp\_opac} \citep{Birnstiel2018a}
    \item \texttt{NumPy} \citep{Harris2020}
    \item \texttt{Matplotlib} \citep{Hunter2007}
    \item \texttt{Astropy} \citep{TheAstropyCollaboration2022}
\end{itemize}

\bibliography{Literature,Literature_Phil}

@article{Baronett2026,
    title = {{A Framework to Model Stellar Irradiated Disks with Frequency-dependent Absorption and Scattering Opacities in Athena++}},
    year = {2026},
    journal = {eprint arXiv:2606.08859},
    author = {Baronett, Stanley A. and Jiang, Yan-Fei and Zhu, Zhaohuan and Zhang, Shangjia and Armitage, Philip J.},
    month = {6},
    pages = {arXiv:2606.08859},
    publisher = {Hydrodynamics},
    url = {http://arxiv.org/abs/2606.08859},
    doi = {10.48550/ARXIV.2606.08859},
    arxivId = {2606.08859}
}

@article{Villenave2022,
    title = {{A Highly Settled Disk around Oph163131}},
    year = {2022},
    journal = {The Astrophysical Journal},
    author = {Villenave, M. and Stapelfeldt, K. R. and Duch{\^{e}}ne, G. and M{\'{e}}nard, F. and Lambrechts, M. and Sierra, A. and Flores, C. and Dent, W. R. F. and Wolff, S. and Ribas, Á. and Benisty, M. and Cuello, N. and Pinte, C.},
    number = {1},
    pages = {11},
    volume = {930},
    url = {https://doi.org/10.3847/1538-4357/ac5fae},
    doi = {10.3847/1538-4357/ac5fae},
    issn = {0004-637X},
    arxivId = {2204.00640}
}

@article{Huang2022,
    title = {{A Multifluid Dust Module in Athena++: Algorithms and Numerical Tests}},
    year = {2022},
    journal = {The Astrophysical Journal Supplement Series},
    author = {Huang, Pinghui and Bai, Xue-Ning},
    number = {1},
    pages = {11},
    volume = {262},
    url = {https://doi.org/10.3847/1538-4365/ac76cb},
    doi = {10.3847/1538-4365/ac76cb},
    issn = {0067-0049},
    arxivId = {2206.01023}
}

@article{Fukuhara2024,
    title = {{A self-consistent model for dust settling and the vertical shear instability in protoplanetary disks}},
    year = {2024},
    journal = {Publications of the Astronomical Society of Japan},
    author = {Fukuhara, Yuya and Okuzumi, Satoshi},
    month = {5},
    url = {http://arxiv.org/abs/2404.15780},
    doi = {10.1093/PASJ/PSAE042},
    issn = {0004-6264},
    arxivId = {2404.15780}
}

@article{Lin2017,
    title = {{A Thermodynamic View of Dusty Protoplanetary Disks}},
    year = {2017},
    journal = {The Astrophysical Journal},
    author = {Lin, Min Kai and Youdin, Andrew N},
    pages = {129},
    volume = {849},
    url = {https://doi.org/10.3847/1538-4357/aa92cd},
    doi = {10.3847/1538-4357/aa92cd}
}

@article{Jiang2021,
    title = {{An Implicit Finite Volume Scheme to Solve the Time-dependent Radiation Transport Equation Based on Discrete Ordinates}},
    year = {2021},
    journal = {The Astrophysical Journal Supplement Series},
    author = {Jiang, Yan-Fei},
    number = {2},
    month = {4},
    pages = {49},
    volume = {253},
    publisher = {American Astronomical Society},
    url = {https://ui.adsabs.harvard.edu/abs/2021ApJS..253...49J/abstract},
    doi = {10.3847/1538-4365/abe303},
    issn = {0067-0049},
    arxivId = {2102.02212}
}

@article{Holton2012,
    title = {{An introduction to dynamic meteorology: Fifth edition}},
    year = {2012},
    journal = {An Introduction to Dynamic Meteorology: Fifth Edition},
    author = {Holton, James R. and Hakim, Gregory J.},
    month = {8},
    pages = {1--532},
    volume = {9780123848},
    publisher = {Academic Press},
    url = {http://www.sciencedirect.com:5070/book/9780123848666/an-introduction-to-dynamic-meteorology},
    isbn = {9780123848666},
    doi = {10.1016/C2009-0-63394-8}
}

@article{Harris2020,
    title = {{Array programming with NumPy}},
    year = {2020},
    journal = {Nature},
    author = {Harris, Charles R. and Millman, K. Jarrod and van der Walt, Stéfan J. and Gommers, Ralf and Virtanen, Pauli and Cournapeau, David and Wieser, Eric and Taylor, Julian and Berg, Sebastian and Smith, Nathaniel J. and Kern, Robert and Picus, Matti and Hoyer, Stephan and van Kerkwijk, Marten H. and Brett, Matthew and Haldane, Allan and del R{\'{i}}o, Jaime Fernández and Wiebe, Mark and Peterson, Pearu and G{\'{e}}rard-Marchant, Pierre and Sheppard, Kevin and Reddy, Tyler and Weckesser, Warren and Abbasi, Hameer and Gohlke, Christoph and Oliphant, Travis E.},
    number = {7825},
    month = {9},
    pages = {357--362},
    volume = {585},
    publisher = {Nature Publishing Group},
    url = {https://www.nature.com/articles/s41586-020-2649-2},
    doi = {10.1038/s41586-020-2649-2},
    issn = {14764687},
    pmid = {32939066},
    arxivId = {2006.10256}
}

@article{Flock2013,
    title = {{Astronomy Astrophysics Radiation magnetohydrodynamics in global simulations of protoplanetary discs}},
    year = {2013},
    journal = {Astronomy and Astrophysics},
    author = {Flock, M and Fromang, S and Gonz{\'{a}}lez, M and Commer{\c{c}}on, B},
    pages = {43},
    volume = {560},
    url = {www.ita.uni-heidelberg.de/~dullemond/software/},
    doi = {10.1051/0004-6361/201322451}
}

@article{Petersen2007b,
    title = {{Baroclinic Vorticity Production in Protoplanetary Disks. I. Vortex Formation}},
    year = {2007},
    journal = {The Astrophysical Journal},
    author = {Petersen, Mark R. and Julien, Keith and Stewart, Glen R.},
    number = {2},
    month = {4},
    pages = {1236--1251},
    volume = {658},
    publisher = {American Astronomical Society},
    url = {http://arxiv.org/abs/astro-ph/0611528 http://dx.doi.org/10.1086/511513 https://ui.adsabs.harvard.edu/abs/2007ApJ...658.1252P/abstract https://iopscience.iop.org/article/10.1086/511513 https://iopscience.iop.org/article/10.1086/511513/meta},
    doi = {10.1086/511513},
    issn = {0004-637X},
    arxivId = {astro-ph/0611528}
}

@article{Lin2015,
    title = {{Cooling requirements for the vertical shear instability in protoplanetary disks}},
    year = {2015},
    journal = {Astrophysical Journal},
    author = {Lin, Min Kai and Youdin, Andrew N.},
    number = {1},
    volume = {811},
    doi = {10.1088/0004-637X/811/1/17},
    issn = {15384357},
    arxivId = {1505.02163}
}

@article{Sauter1926,
    title = {{Die Gr{\"{o}}ssenbestimmung der im Gemischnebel von Verbrennungskraftmaschinen vorhandenen Brennstoffteilchen:(Mitteilung aus d. Labor. f. techn. Physik d. Techn. Hochsch. M{\"{u}}nchen)}},
    year = {1926},
    journal = {VDI-Verlag, Forschungsarbeiten auf dem Gebiete des Ingenieurwesens},
    author = {Sauter, J.},
    volume = {279},
    url = {https://books.google.de/books?id=5pcinQEACAAJ&redir_esc=y}
}

@misc{Birnstiel2018a,
    title = {{dsharp{\textbackslash}{\_}opac: Revised Release of Package}},
    year = {2018},
    booktitle = {Zenodo},
    author = {Birnstiel, Til},
    month = {11},
    url = {https://ui.adsabs.harvard.edu/abs/2018zndo...1495277B/abstract},
    doi = {10.5281/ZENODO.1495277}
}

@article{Pfeil2023,
    title = {{Dust Coagulation Reconciles Protoplanetary Disk Observations with the Vertical Shear Instability. I. Dust Coagulation and the VSI Dead Zone}},
    year = {2023},
    journal = {The Astrophysical Journal},
    author = {Pfeil, Thomas and Birnstiel, Tilman and Klahr, Hubert},
    number = {2},
    month = {10},
    pages = {121},
    volume = {959},
    url = {http://arxiv.org/abs/2310.07332},
    doi = {10.3847/1538-4357/ad00af},
    issn = {0004-637X},
    arxivId = {2310.07332}
}

@article{Morrissey2026,
    title = {{Dust Collisions in Protoplanetary Disks: Atomic Simulations of the Surface Free Energy}},
    year = {2026},
    journal = {Astronomical Journal},
    author = {Morrissey, L. S. and Ebel, D. S. and Eriksson, L. E.J. and Georgiou, A. and Huang, Z. and Mac Low, M. M. and Pfeil, T.},
    number = {1},
    month = {1},
    pages = {42},
    volume = {171},
    publisher = {American Astronomical Society},
    url = {https://ui.adsabs.harvard.edu/abs/2026AJ....171...42M/abstract},
    doi = {10.3847/1538-3881/ae1976},
    issn = {15383881}
}

@article{Blum2018,
    title = {{Dust Evolution in Protoplanetary Discs and the Formation of Planetesimals: What Have We Learned from Laboratory Experiments?}},
    year = {2018},
    journal = {Space Science Reviews},
    author = {Blum, Jürgen},
    number = {2},
    month = {2},
    pages = {1--19},
    volume = {214},
    publisher = {Springer},
    url = {https://link.springer.com/article/10.1007/s11214-018-0486-5},
    doi = {10.1007/s11214-018-0486-5},
    issn = {15729672},
    arxivId = {1802.00221}
}

@article{Lin2019,
    title = {{Dust settling against hydrodynamic turbulence in protoplanetary discs}},
    year = {2019},
    journal = {Monthly Notices of the Royal Astronomical Society},
    author = {Lin, Min Kai},
    number = {4},
    month = {6},
    pages = {5221--5234},
    volume = {485},
    publisher = {Oxford Academic},
    url = {https://dx.doi.org/10.1093/mnras/stz701},
    doi = {10.1093/MNRAS/STZ701},
    issn = {0035-8711},
    arxivId = {1903.03620}
}

@article{Stammler2022,
    title = {{DustPy: A Python Package for Dust Evolution in Protoplanetary Disks}},
    year = {2022},
    journal = {The Astrophysical Journal},
    author = {Stammler, Sebastian M. and Birnstiel, Tilman},
    number = {1},
    pages = {35},
    volume = {935},
    url = {https://doi.org/10.3847/1538-4357/ac7d58},
    doi = {10.3847/1538-4357/ac7d58},
    issn = {0004-637X},
    arxivId = {2207.00322}
}

@article{Fukuhara2021,
    title = {{Effects of Dust Evolution on the Vertical Shear Instability in the Outer Regions of Protoplanetary Disks}},
    year = {2021},
    journal = {The Astrophysical Journal},
    author = {Fukuhara, Yuya and Okuzumi, Satoshi and Ono, Tomohiro},
    number = {2},
    pages = {132},
    volume = {914},
    url = {https://doi.org/10.3847/1538-4357/abfe5c},
    doi = {10.3847/1538-4357/abfe5c},
    issn = {0004-637X},
    arxivId = {2105.02403}
}

@article{Malygin2017,
    title = {{Efficiency of thermal relaxation by radiative processes in protoplanetary discs: Constraints on hydrodynamic turbulence}},
    year = {2017},
    journal = {Astronomy and Astrophysics},
    author = {Malygin, M. G. and Klahr, H. and Semenov, D. and Henning, Th and Dullemond, C. P.},
    pages = {30},
    volume = {605},
    doi = {10.1051/0004-6361/201629933},
    issn = {14320746},
    arxivId = {1704.06786}
}

@article{Barraza-Alfaro2025,
    title = {{exoALMA. XVI. Predicting Signatures of Large-scale Turbulence in Protoplanetary Disks}},
    year = {2025},
    journal = {The Astrophysical Journal Letters},
    author = {Barraza-Alfaro, Marcelo and Flock, Mario and B{\'{e}}thune, William and Teague, Richard and Bae, Jaehan and Benisty, Myriam and Cataldi, Gianni and Curone, Pietro and Czekala, Ian and Facchini, Stefano and Fasano, Daniele and Fukagawa, Misato and Galloway-Sprietsma, Maria and Garg, Himanshi and Hall, Cassandra and Huang, Jane and Ilee, John D. and Izquierdo, Andrés F. and Kanagawa, Kazuhiro and Koch, Eric W. and Lesur, Geoffroy and Longarini, Cristiano and Loomis, Ryan A. and Orihara, Ryuta and Pinte, Christophe and Price, Daniel J. and Rosotti, Giovanni and Stadler, Jochen and Wafflard-Fernandez, Gaylor and Winter, Andrew J. and W{\"{o}}lfer, Lisa and Yen, Hsi-Wei and Yoshida, Tomohiro C. and Zawadzki, Brianna},
    number = {1},
    month = {5},
    pages = {L21},
    volume = {984},
    publisher = {American Astronomical Society},
    url = {https://ui.adsabs.harvard.edu/abs/2025ApJ...984L..21B/abstract},
    doi = {10.3847/2041-8213/adc42d},
    issn = {2041-8205},
    arxivId = {2504.19853}
}

@article{Flock2020,
    title = {{Gas and Dust Dynamics in Starlight-heated Protoplanetary Disks}},
    year = {2020},
    journal = {The Astrophysical Journal},
    author = {Flock, Mario and Turner, Neal J. and Nelson, Richard P. and Lyra, Wladimir and Manger, Natascha and Klahr, Hubert},
    number = {2},
    pages = {155},
    volume = {897},
    url = {https://doi.org/10.3847/1538-4357/ab9641},
    doi = {10.3847/1538-4357/ab9641},
    issn = {15384357},
    arxivId = {2005.11974}
}

@article{Flores-Rivera2026,
    title = {{Gas dynamics around dust asymmetries in turbulent disks}},
    year = {2026},
    journal = {Astronomy {\&} Astrophysics},
    author = {Flores-Rivera, Lizxandra and Manger, Natascha and Lambrechts, Michiel and Flock, Mario and Lorek, Sebastian and Johansen, Anders and Klahr, Hubert},
    month = {6},
    pages = {A124},
    volume = {710},
    publisher = {EDP Sciences},
    url = {https://ui.adsabs.harvard.edu/abs/2026A&A...710A.124F/abstract},
    doi = {10.1051/0004-6361/202557487},
    issn = {0004-6361},
    arxivId = {2603.24184}
}

@article{growpacity,
    title = {{growpacity: A computationally efficient dust opacity model suitable for coagulation models}},
    year = {2026},
    journal = {ascl},
    author = {Ziampras, Alexandros and Birnstiel, Tilman},
    pages = {ascl:2603.020},
    url = {https://ui.adsabs.harvard.edu/abs/2026ascl.soft03020Z/abstract}
}

@article{Lesur2025,
    title = {{High-resolution models of the vertical shear instability}},
    year = {2025},
    journal = {Astronomy and Astrophysics},
    author = {Lesur, G. and Latter, H. and Ogilvie, G. I.},
    volume = {703},
    url = {http://arxiv.org/abs/2508.20839},
    doi = {10.1051/0004-6361/202555944},
    issn = {14320746},
    arxivId = {2508.20839}
}

@article{Manger2021,
    title = {{High-resolution parameter study of the vertical shear instability-II: Dependence on temperature gradient and cooling time}},
    year = {2021},
    journal = {Monthly Notices of the Royal Astronomical Society},
    author = {Manger, Natascha and Pfeil, Thomas and Klahr, Hubert},
    number = {4},
    pages = {5402--5409},
    volume = {508},
    url = {https://doi.org/10.1093/mnras/stab2599},
    doi = {10.1093/mnras/stab2599},
    issn = {13652966},
    arxivId = {2109.01649}
}

@incollection{Lesur2022,
    title = {{Hydro-, Magnetohydro-, and Dust-Gas Dynamics of Protoplanetary Disks}},
    year = {2023},
    booktitle = {Protostars and Planets VII},
    author = {Lesur, G. and Flock, M. and Ercolano, B. and Lin, M. -K. and Yang, C. -C. and Barranco, J. A. and Benitez-Llambay, P. and Goodman, J. and Johansen, A. and Klahr, H. and Laibe, G. and Lyra, W. and Marcus, P. S. and Nelson, R. P. and Squire, J. and Simon, J. B. and Turner, N. J. and Umurhan, O. M. and Youdin, A. N. and Lesur, G. and Flock, M. and Ercolano, B. and Lin, M. -K. and Yang, C. -C. and Barranco, J. A. and Benitez-Llambay, P. and Goodman, J. and Johansen, A. and Klahr, H. and Laibe, G. and Lyra, W. and Marcus, P. S. and Nelson, R. P. and Squire, J. and Simon, J. B. and Turner, N. J. and Umurhan, O. M. and Youdin, A. N. and Ercolano, B. and Lin, M. -K. and Yang, C. -C. and Barranco, J. A. and Benitez-Llambay, P. and Goodman, J. and Johansen, A. and Klahr, H. and Laibe, G. and Lyra, W. and Marcus, P. S. and Nelson, R. P. and Squire, J. and Simon, J. B. and Turner, N. J. and Umurhan, O. M. and Youdin, A. N. and Lesur, G. and Flock, M. and Ercolano, B. and Lin, M. -K. and Yang, C. -C. and Barranco, J. A. and Benitez-Llambay, P. and Goodman, J. and Johansen, A. and Klahr, H. and Laibe, G. and Lyra, W. and Marcus, P. S. and Nelson, R. P. and Squire, J. and Simon, J. B. and Turner, N. J. and Umurhan, O. M. and Youdin, A. N.},
    editor = {Inutsuka, Shu-ichiro and Aikawa, Yuri and Muto, Takayuki and Tomida, Kengo and Tamura, Motohide},
    month = {3},
    pages = {465--500},
    volume = {534},
    publisher = {ASPSC},
    url = {https://ui.adsabs.harvard.edu/abs/2023ASPC..534..465L/abstract http://arxiv.org/abs/2203.09821},
    address = {San Francisco},
    issn = {1050-3390},
    arxivId = {2203.09821}
}

@article{Fukuhara2025,
    title = {{Hydrodynamical simulations of the vertical shear instability with dynamic dust and cooling rates in protoplanetary disks}},
    year = {2025},
    journal = {Astronomy and Astrophysics},
    author = {Fukuhara, Y. and Flock, M. and Okuzumi, S. and Tominaga, R. T.},
    month = {7},
    pages = {A72},
    volume = {701},
    publisher = {EDP Sciences},
    url = {http://arxiv.org/abs/2507.11156 http://dx.doi.org/10.1051/0004-6361/202555624},
    doi = {10.1051/0004-6361/202555624},
    issn = {14320746},
    arxivId = {2507.11156}
}

@article{Barraza-Alfaro2024,
    title = {{Kinematic signatures of planet–disk interactions in vertical shear instability-turbulent protoplanetary disks}},
    year = {2024},
    journal = {Astronomy and Astrophysics},
    author = {Barraza-Alfaro, Marcelo and Flock, Mario and Henning, Thomas},
    volume = {683},
    url = {https://doi.org/10.1051/0004-6361/202347726},
    doi = {10.1051/0004-6361/202347726},
    issn = {14320746}
}

@article{Nelson2013,
    title = {{Linear and non-linear evolution of the vertical shear instability in accretion discs}},
    year = {2013},
    journal = {Monthly Notices of the Royal Astronomical Society},
    author = {Nelson, Richard P. and Gressel, Oliver and Umurhan, Orkan M.},
    number = {3},
    pages = {2610--2632},
    volume = {435},
    url = {https://academic.oup.com/mnras/article/435/3/2610/1031890},
    doi = {10.1093/mnras/stt1475},
    issn = {13652966},
    arxivId = {1209.2753}
}

@article{Urpin1998,
    title = {{Magnetic and vertical shear instabilities in accretion discs}},
    year = {1998},
    journal = {Monthly Notices of the Royal Astronomical Society},
    author = {Urpin, V. and Brandenburg, A.},
    number = {3},
    pages = {399--406},
    volume = {294},
    doi = {10.1111/j.1365-8711.1998.01118.x},
    issn = {00358711}
}

@article{Pfeil2019,
    title = {{Mapping the Conditions for Hydrodynamic Instability on Steady-State Accretion Models of Protoplanetary Disks}},
    year = {2019},
    journal = {The Astrophysical Journal},
    author = {Pfeil, Thomas and Klahr, Hubert},
    number = {2},
    pages = {150},
    volume = {871},
    url = {https://doi.org/10.3847/1538-4357/aaf962},
    doi = {10.3847/1538-4357/aaf962},
    issn = {15384357},
    arxivId = {1808.10344}
}

@article{Hunter2007,
    title = {{Matplotlib: A 2D graphics environment}},
    year = {2007},
    journal = {Computing in Science and Engineering},
    author = {Hunter, John D.},
    number = {3},
    pages = {90--95},
    volume = {9},
    publisher = {IEEE Computer Society},
    doi = {10.1109/MCSE.2007.55},
    issn = {15219615}
}

@article{Baraffe2015,
    title = {{New evolutionary models for pre-main sequence and main sequence low-mass stars down to the hydrogen-burning limit}},
    year = {2015},
    journal = {Astronomy and Astrophysics},
    author = {Baraffe, Isabelle and Homeier, Derek and Allard, France and Chabrier, Gilles},
    volume = {577},
    doi = {10.1051/0004-6361/201425481},
    issn = {14320746}
}

@article{Fuksman2022,
    title = {{No Self-shadowing Instability in 2D Radiation Hydrodynamical Models of Irradiated Protoplanetary Disks}},
    year = {2022},
    journal = {The Astrophysical Journal},
    author = {Melon Fuksman, Julio David and Klahr, Hubert},
    number = {1},
    month = {9},
    pages = {16},
    volume = {936},
    publisher = {American Astronomical Society},
    url = {https://ui.adsabs.harvard.edu/abs/2022ApJ...936...16M/abstract},
    doi = {10.3847/1538-4357/ac7fee},
    issn = {0004-637X},
    arxivId = {2207.05106}
}

@article{Stoll2016,
    title = {{Particle dynamics in discs with turbulence generated by the vertical shear instability}},
    year = {2016},
    journal = {Astronomy and Astrophysics},
    author = {Stoll, Moritz H.R. R and Kley, Wilhelm},
    pages = {57},
    volume = {594},
    doi = {10.1051/0004-6361/201527716},
    issn = {14320746},
    arxivId = {1607.02322}
}

@article{Jiang2025,
    title = {{Puffed-up Inner Rings and Razor-thin Outer Rings in Structured Protoplanetary Disks}},
    year = {2025},
    journal = {The Astrophysical Journal},
    author = {Jiang, Haochang and Long, Feng and Mac{\'{i}}as, Enrique and Benisty, Myriam and Doi 土井, Kiyoaki 聖明 and Dullemond, Cornelis P. and Loomis, Ryan A. and Pascucci, Ilaria and P{\'{e}}rez, Sebastián and Zhang 张, Shangjia 尚嘉 and Zhu 朱, Zhaohuan 照寰},
    number = {2},
    month = {11},
    pages = {166},
    volume = {993},
    publisher = {American Astronomical Society},
    url = {https://ui.adsabs.harvard.edu/abs/2025ApJ...993..166J/abstract},
    doi = {10.3847/1538-4357/ae089e},
    issn = {0004-637X},
    arxivId = {arXiv:2509.13122}
}

@article{Dullemond2012,
    title = {{RADMC-3D: A multi-purpose radiative transfer tool}},
    year = {2012},
    journal = {ascl},
    author = {Dullemond, C. P. and Juhasz, A. and Pohl, A. and Sereshti, F. and Shetty, R. and Peters, T. and Commercon, B. and Flock, M. and Dullemond, C. P. and Juhasz, A. and Pohl, A. and Sereshti, F. and Shetty, R. and Peters, T. and Commercon, B. and Flock, M.},
    pages = {ascl:1202.015},
    url = {https://ui.adsabs.harvard.edu/abs/2012ascl.soft02015D/abstract}
}

@article{Dullemond2022,
    title = {{Razor-thin dust layers in protoplanetary disks: Limits on the vertical shear instability}},
    year = {2022},
    journal = {Astronomy and Astrophysics},
    author = {Dullemond, C. P. and Ziampras, A. and Ostertag, D. and Dominik, C.},
    pages = {105},
    volume = {668},
    url = {https://doi.org/10.1051/0004-6361/202244218},
    doi = {10.1051/0004-6361/202244218},
    issn = {14320746},
    arxivId = {2210.13413}
}

@article{Youdin2005,
    title = {{Streaming Instabilities in Protoplanetary Disks}},
    year = {2005},
    journal = {The Astrophysical Journal},
    author = {Youdin, Andrew N. and Goodman, Jeremy},
    number = {1},
    pages = {459--469},
    volume = {620},
    doi = {10.1086/426895},
    issn = {0004-637X},
    arxivId = {astro-ph/0409263}
}

@article{TheAstropyCollaboration2022,
    title = {{The Astropy Project: Sustaining and Growing a Community-oriented Open-source Project and the Latest Major Release (v5.0) of the Core Package*}},
    year = {2022},
    journal = {The Astrophysical Journal},
    author = {{The Astropy Collaboration} and Price-Whelan, Adrian M. and Lim, Pey Lian and Earl, Nicholas and Starkman, Nathaniel and Bradley, Larry and Shupe, David L. and Patil, Aarya A. and Corrales, Lia and Brasseur, C. E. and N{\"{o}}the, Maximilian and Donath, Axel and Tollerud, Erik and Morris, Brett M. and Ginsburg, Adam and Vaher, Eero and Weaver, Benjamin A. and Tocknell, James and Jamieson, William and van Kerkwijk, Marten H. and Robitaille, Thomas P. and Merry, Bruce and Bachetti, Matteo and G{\"{u}}nther, H. Moritz and Aldcroft, Thomas L. and Alvarado-Montes, Jaime A. and Archibald, Anne M. and B{\'{o}}di, Attila and Bapat, Shreyas and Barentsen, Geert and Baz{\'{a}}n, Juanjo and Biswas, Manish and Boquien, Médéric and Burke, D. J. and Cara, Daria and Cara, Mihai and Conroy, Kyle E and Conseil, Simon and Craig, Matthew W. and Cross, Robert M. and Cruz, Kelle L. and D’Eugenio, Francesco and Dencheva, Nadia and Devillepoix, Hadrien A. R. and Dietrich, Jörg P. and Eigenbrot, Arthur Davis and Erben, Thomas and Ferreira, Leonardo and Foreman-Mackey, Daniel and Fox, Ryan and Freij, Nabil and Garg, Suyog and Geda, Robel and Glattly, Lauren and Gondhalekar, Yash and Gordon, Karl D. and Grant, David and Greenfield, Perry and Groener, Austen M. and Guest, Steve and Gurovich, Sebastian and Handberg, Rasmus and Hart, Akeem and Hatfield-Dodds, Zac and Homeier, Derek and Hosseinzadeh, Griffin and Jenness, Tim and Jones, Craig K. and Joseph, Prajwel and Kalmbach, J. Bryce and Karamehmetoglu, Emir and Ka{\l}uszy{\'{n}}ski, Mikołaj and Kelley, Michael S. P. and Kern, Nicholas and Kerzendorf, Wolfgang E. and Koch, Eric W. and Kulumani, Shankar and Lee, Antony and Ly, Chun and Ma, Zhiyuan and MacBride, Conor and Maljaars, Jakob M. and Muna, Demitri and Murphy, N. A. and Norman, Henrik and O’Steen, Richard and Oman, Kyle A. and Pacifici, Camilla and Pascual, Sergio and Pascual-Granado, J. and Patil, Rohit R. and Perren, Gabriel I and Pickering, Timothy E. and Rastogi, Tanuj and Roulston, Benjamin R. and Ryan, Daniel F and Rykoff, Eli S. and Sabater, Jose and Sakurikar, Parikshit and Salgado, Jesús and Sanghi, Aniket and Saunders, Nicholas and Savchenko, Volodymyr and Schwardt, Ludwig and Seifert-Eckert, Michael and Shih, Albert Y. and Jain, Anany Shrey and Shukla, Gyanendra and Sick, Jonathan and Simpson, Chris and Singanamalla, Sudheesh and Singer, Leo P. and Singhal, Jaladh and Sinha, Manodeep and Sip{\H{o}}cz, Brigitta M. and Spitler, Lee R. and Stansby, David and Streicher, Ole and {\v{S}}umak, Jani and Swinbank, John D. and Taranu, Dan S. and Tewary, Nikita and Tremblay, Grant R. and Val-Borro, Miguel de and Van Kooten, Samuel J. and Vasovi{\'{c}}, Zlatan and Verma, Shresth and de Miranda Cardoso, José Vinícius and Williams, Peter K. G. and Wilson, Tom J. and Winkel, Benjamin and Wood-Vasey, W. M. and Xue, Rui and Yoachim, Peter and Zhang, Chen and Zonca, Andrea},
    number = {2},
    month = {8},
    pages = {167},
    volume = {935},
    publisher = {American Astronomical Society},
    url = {https://ui.adsabs.harvard.edu/abs/2022ApJ...935..167A/abstract},
    doi = {10.3847/1538-4357/ac7c74},
    issn = {0004-637X},
    arxivId = {2206.14220}
}

@article{Stone2020,
    title = {{The Athena++ Adaptive Mesh Refinement Framework: Design and Magnetohydrodynamic Solvers}},
    year = {2020},
    journal = {The Astrophysical Journal Supplement Series},
    author = {Stone, James M. and Tomida, Kengo and White, Christopher J. and Felker, Kyle G.},
    number = {1},
    pages = {4},
    volume = {249},
    url = {https://doi.org/10.3847/1538-4365/ab929b},
    doi = {10.3847/1538-4365/ab929b},
    issn = {00670049},
    arxivId = {2005.06651}
}

@article{Schafer2020,
    title = {{The coexistence of the streaming instability and the vertical shear instability in protoplanetary disks}},
    year = {2020},
    journal = {Astronomy and Astrophysics},
    author = {Sch{\"{a}}fer, Urs and Johansen, Anders and Banerjee, Robi},
    volume = {635},
    url = {https://doi.org/10.1051/0004-6361/201937371},
    doi = {10.1051/0004-6361/201937371},
    issn = {14320746},
    arxivId = {2002.07185}
}

@article{Schafer2022,
    title = {{The coexistence of the streaming instability and the vertical shear instability in protoplanetary disks: Planetesimal formation thresholds explored in two-dimensional global models}},
    year = {2022},
    journal = {Astronomy and Astrophysics},
    author = {Sch{\"{a}}fer, Urs and Johansen, Anders},
    pages = {A98},
    volume = {666},
    url = {https://doi.org/10.1051/0004-6361/202243655},
    doi = {10.1051/0004-6361/202243655},
    issn = {14320746}
}

@article{Birnstiel2018,
    title = {{The Disk Substructures at High Angular Resolution Project (DSHARP). V. Interpreting ALMA Maps of Protoplanetary Disks in Terms of a Dust Model}},
    year = {2018},
    journal = {The Astrophysical Journal},
    author = {Birnstiel, Tilman and Dullemond, Cornelis P. and Zhu, Zhaohuan and Andrews, Sean M. and Bai, Xue-Ning and Wilner, David J. and Carpenter, John M. and Huang, Jane and Isella, Andrea and Benisty, Myriam and P{\'{e}}rez, Laura M. and Zhang, Shangjia},
    number = {2},
    pages = {L45},
    volume = {869},
    url = {https://doi.org/10.3847/2041-8213/aaf743},
    doi = {10.3847/2041-8213/aaf743},
    issn = {20418213},
    arxivId = {1812.04043}
}

@article{LyndenBell1974,
    title = {{The Evolution of Viscous Discs and the Origin of the Nebular Variables}},
    year = {1974},
    journal = {Monthly Notices of the Royal Astronomical Society},
    author = {Lynden-Bell, D. and Pringle, J. E.},
    number = {3},
    pages = {603--637},
    volume = {168},
    doi = {10.1093/mnras/168.3.603},
    issn = {0035-8711}
}

@article{Huang2025a,
    title = {{The Interplay between Dust Dynamics and Turbulence Induced by the Vertical Shear Instability}},
    year = {2025},
    journal = {The Astrophysical Journal},
    author = {Huang, Pinghui and Bai, Xue-Ning},
    number = {1},
    month = {6},
    pages = {76},
    volume = {986},
    publisher = {American Astronomical Society},
    url = {https://ui.adsabs.harvard.edu/abs/2025ApJ...986...76H/abstract},
    doi = {10.3847/1538-4357/add345},
    issn = {0004-637X},
    arxivId = {arXiv:2503.01656}
}

@article{Zsom2011,
    title = {{The outcome of protoplanetary dust growth: Pebbles, boulders, or planetesimals?: III. Sedimentation driven coagulation inside the snowline}},
    year = {2011},
    journal = {Astronomy and Astrophysics},
    author = {Zsom, A. and Ormel, C. W. and Dullemond, C. P. and Henning, T.},
    month = {10},
    pages = {A73},
    volume = {534},
    publisher = {EDP Sciences},
    url = {https://www.aanda.org/articles/aa/full_html/2011/10/aa16515-11/aa16515-11.html https://www.aanda.org/articles/aa/abs/2011/10/aa16515-11/aa16515-11.html},
    doi = {10.1051/0004-6361/201116515},
    issn = {00046361},
    arxivId = {1107.5198}
}

@article{Pfeil2021,
    title = {{The Sandwich Mode for Vertical Shear Instability in Protoplanetary Disks}},
    year = {2021},
    journal = {The Astrophysical Journal},
    author = {Pfeil, Thomas and Klahr, Hubert},
    number = {2},
    pages = {130},
    volume = {915},
    url = {https://doi.org/10.3847/1538-4357/ac0054},
    doi = {10.3847/1538-4357/ac0054},
    issn = {0004-637X},
    arxivId = {2008.11195}
}

@article{Mathis1977,
    title = {{The size distribution of interstellar grains}},
    year = {1977},
    journal = {The Astrophysical Journal},
    author = {Mathis, J. S. and Rumpl, W. and Nordsieck, K. H.},
    pages = {425},
    volume = {217},
    doi = {10.1086/155591},
    issn = {0004-637X}
}

@article{Blanco2021,
    title = {{The Substructures in Disks Undergoing Vertical Shear Instability. II. Observational Predictions for the Dust Continuum}},
    year = {2021},
    journal = {The Astrophysical Journal},
    author = {Blanco, Diana and Ricci, Luca and Flock, Mario and Turner, Neal},
    number = {2},
    pages = {70},
    volume = {920},
    url = {https://doi.org/10.3847/1538-4357/ac15fa},
    doi = {10.3847/1538-4357/ac15fa},
    issn = {0004-637X},
    arxivId = {2108.00907}
}

@article{Ogilvie2025,
    title = {{The vertical shear instability in protoplanetary discs as an outwardly travelling wave - I. Linear theory}},
    year = {2025},
    journal = {Monthly Notices of the Royal Astronomical Society},
    author = {Ogilvie, Gordon I. and Latter, Henrik N. and Lesur, Geoffroy},
    number = {4},
    month = {3},
    pages = {3349--3365},
    volume = {537},
    publisher = {Oxford University Press},
    doi = {10.1093/mnras/staf154},
    issn = {13652966},
    arxivId = {2501.13715}
}

@article{Klahr2026,
    title = {{Thermal Baroclinic Instabilities in Accretion Disks. II. Numerical Experiments for the Goldreich–Schubert–Fricke Instability and the Convective Overstability in Disks around Young Stars}},
    year = {2026},
    journal = {The Astrophysical Journal},
    author = {Klahr, Hubert and Baehr, Hans and Melon Fuksman, Julio David},
    number = {2},
    pages = {211},
    volume = {998},
    url = {https://doi.org/10.3847/1538-4357/ae0f9d},
    doi = {10.3847/1538-4357/ae0f9d},
    issn = {0004-637X}
}

@article{Zhang2024,
    title = {{Thermal Structure Determines Kinematics: Vertical Shear Instability in Stellar Irradiated Protoplanetary Disks}},
    year = {2024},
    journal = {The Astrophysical Journal},
    author = {Zhang, Shangjia and Zhu, Zhaohuan and Jiang, Yan-Fei},
    number = {1},
    month = {4},
    pages = {29},
    volume = {968},
    url = {https://arxiv.org/abs/2404.05608v2},
    doi = {10.3847/1538-4357/ad4109},
    issn = {0004-637X},
    arxivId = {2404.05608}
}

@article{Pfeil2024b,
    title = {{TriPoD: Tri-Population size distributions for Dust evolution. Coagulation in vertically integrated hydrodynamic simulations of protoplanetary disks}},
    year = {2024},
    journal = {Astronomy {\&} Astrophysics},
    author = {Pfeil, Thomas and Birnstiel, Til and Klahr, Hubert},
    month = {10},
    pages = {A45},
    volume = {691},
    publisher = {EDP Sciences},
    url = {http://arxiv.org/abs/2409.03816 http://dx.doi.org/10.1051/0004-6361/202449337},
    doi = {10.1051/0004-6361/202449337},
    arxivId = {2409.03816}
}

@article{Klahr2003,
    title = {{Turbulence in Accretion Disks: Vorticity Generation and Angular Momentum Transport via the Global Baroclinic Instability}},
    year = {2003},
    journal = {The Astrophysical Journal},
    author = {Klahr, H. H. and Bodenheimer, P.},
    number = {2},
    month = {1},
    pages = {869--892},
    volume = {582},
    publisher = {American Astronomical Society},
    url = {https://iopscience.iop.org/article/10.1086/344743 https://iopscience.iop.org/article/10.1086/344743/meta},
    doi = {10.1086/344743},
    issn = {0004-637X},
    arxivId = {astro-ph/0211629}
}

@article{Villenave2025a,
    title = {{Turbulence in protoplanetary disks: A systematic analysis of dust settling in 33 disks}},
    year = {2025},
    journal = {Astronomy and Astrophysics},
    author = {Villenave, Marion and Rosotti, Giovanni P. and Lambrechts, Michiel and Ziampras, Alexandros and Pinte, Christophe and M{\'{e}}nard, François and Stapelfeldt, Karl R. and Duch{\^{e}}ne, Gaspard and Baylock, Emily and Doi, Kiyoaki},
    month = {5},
    pages = {A64},
    volume = {697},
    publisher = {EDP Sciences},
    url = {https://ui.adsabs.harvard.edu/abs/2025A&A...697A..64V/abstract},
    doi = {10.1051/0004-6361/202553822},
    issn = {14320746},
    arxivId = {arXiv:2503.05872}
}

@article{Eriksson2026,
    title = {{Turning the knobs on dust evolution: comparing codes, parameters, and their effects on planet formation and disc observables}},
    year = {2026},
    journal = {Monthly Notices of the Royal Astronomical Society},
    author = {Eriksson, Linn E.J. and Pfeil, Thomas and Kaufmann, Nicolas and Vaikundaraman, Vignesh},
    number = {3},
    month = {7},
    pages = {stag982},
    volume = {549},
    publisher = {Oxford University Press},
    url = {https://ui.adsabs.harvard.edu/abs/2026MNRAS.549ag982E/abstract},
    isbn = {982/8694700},
    doi = {10.1093/mnras/stag982},
    issn = {13652966},
    arxivId = {arXiv:2603.22550}
}

@article{Fukuhara2023,
    title = {{Two saturated states of the vertical shear instability in protoplanetary disks with vertically varying cooling times}},
    year = {2023},
    journal = {Publications of the Astronomical Society of Japan},
    author = {Fukuhara, Yuya and Okuzumi, Satoshi and Ono, Tomohiro},
    number = {1},
    pages = {233--249},
    volume = {75},
    url = {https://doi.org/10.1093/pasj/psac107},
    doi = {10.1093/pasj/psac107},
    issn = {2053051X}
}

@article{Flores-Rivera2025,
    title = {{UV processing of icy pebbles in the outer parts of VSI-turbulent disks}},
    year = {2025},
    journal = {Astronomy and Astrophysics},
    author = {Flores-Rivera, Lizxandra and Lambrechts, Michiel and Gavino, Sacha and Lorek, Sebastian and Flock, Mario and Johansen, Anders and Mignone, Andrea},
    month = {12},
    pages = {A281},
    volume = {693},
    publisher = {EDP Sciences},
    url = {http://arxiv.org/abs/2412.01698 http://dx.doi.org/10.1051/0004-6361/202452933},
    doi = {10.1051/0004-6361/202452933},
    issn = {14320746},
    arxivId = {2412.01698}
}

@article{Yun2025a,
    title = {{Vertical Shear Instability in Thermally Stratified Protoplanetary Disks. I. A Linear Stability Analysis}},
    year = {2025},
    journal = {The Astrophysical Journal},
    author = {Yun, Han-Gyeol and Kim, Woong-Tae and Bae, Jaehan and Han, Cheongho},
    number = {1},
    month = {2},
    pages = {14},
    volume = {980},
    publisher = {American Astronomical Society},
    url = {https://ui.adsabs.harvard.edu/abs/2025ApJ...980...14Y/abstract},
    doi = {10.3847/1538-4357/ad9f41},
    issn = {0004-637X},
    arxivId = {arXiv:2412.09924}
}

@article{Yun2025,
    title = {{Vertical Shear Instability in Thermally Stratified Protoplanetary Disks. II. Hydrodynamic Simulations and Observability}},
    year = {2025},
    journal = {The Astrophysical Journal},
    author = {Yun, Han-Gyeol and Kim, Woong-Tae and Bae, Jaehan and Han, Cheongho},
    number = {1},
    month = {2},
    pages = {15},
    volume = {980},
    publisher = {American Astronomical Society},
    url = {https://ui.adsabs.harvard.edu/abs/2025ApJ...980...15Y/abstract},
    doi = {10.3847/1538-4357/ad9f42},
    issn = {0004-637X},
    arxivId = {2412.09930}
}

@article{Pfeil2024,
    title = {{Vertical shear instability with dust evolution and consistent cooling times - On the importance of the initial dust distribution}},
    year = {2024},
    journal = {Astronomy {\&} Astrophysics},
    author = {Pfeil, Thomas and Birnstiel, Til and Klahr, Hubert},
    month = {7},
    pages = {L5},
    volume = {687},
    publisher = {EDP Sciences},
    url = {https://www.aanda.org/articles/aa/full_html/2024/07/aa49323-24/aa49323-24.html https://www.aanda.org/articles/aa/abs/2024/07/aa49323-24/aa49323-24.html},
    doi = {10.1051/0004-6361/202449323},
    issn = {0004-6361},
    arxivId = {astro-ph/0408524}
}

@article{Manger2018,
    title = {{Vortex formation and survival in protoplanetary discs subject to vertical shear instability}},
    year = {2018},
    journal = {Monthly Notices of the Royal Astronomical Society},
    author = {Manger, Natascha and Klahr, Hubert},
    number = {2},
    pages = {2125--2136},
    volume = {480},
    url = {https://academic.oup.com/mnras/article/480/2/2125/5056213},
    doi = {10.1093/MNRAS/STY1909},
    issn = {13652966},
    arxivId = {1807.06492}
}

@article{Richard2016,
    title = {{Vortex formation in protoplanetary discs induced by the vertical shear instability}},
    year = {2016},
    journal = {Monthly Notices of the Royal Astronomical Society},
    author = {Richard, Samuel and Nelson, Richard P. and Umurhan, Orkan M.},
    number = {4},
    pages = {3571--3584},
    volume = {456},
    url = {https://academic.oup.com/mnras/article/456/4/3571/1032344},
    doi = {10.1093/mnras/stv2898},
    issn = {13652966},
    arxivId = {1601.01921}
}

@article{Svanberg2022,
    title = {{Wavelike nature of the vertical shear instability in global protoplanetary discs}},
    year = {2022},
    journal = {Monthly Notices of the Royal Astronomical Society},
    author = {Svanberg, Eleonora and Cui, Can and Latter, Henrik N.},
    number = {3},
    pages = {4581--4587},
    volume = {514},
    url = {https://doi.org/10.1093/mnras/stac1598},
    doi = {10.1093/mnras/stac1598},
    issn = {13652966}
}

@article{Barranco2018,
    title = {{Zombie Vortex Instability. III. Persistence with Nonuniform Stratification and Radiative Damping}},
    year = {2018},
    journal = {The Astrophysical Journal},
    author = {Barranco, Joseph A. and Pei, Suyang and Marcus, Philip S.},
    number = {2},
    pages = {127},
    volume = {869},
    url = {https://doi.org/10.3847/1538-4357/aaec80},
    doi = {10.3847/1538-4357/aaec80},
    issn = {15384357},
    arxivId = {1810.06588}
}

@ARTICLE{Kratter2016,
       author = {{Kratter}, Kaitlin and {Lodato}, Giuseppe},
        title = "{Gravitational Instabilities in Circumstellar Disks}",
      journal = {\araa},
         year = 2016,
        month = sep,
       volume = {54},
        pages = {271-311},
          doi = {10.1146/annurev-astro-081915-023307},
archivePrefix = {arXiv},
       eprint = {1603.01280},
 primaryClass = {astro-ph.SR},
       adsurl = {https://ui.adsabs.harvard.edu/abs/2016ARA&A..54..271K}
}

@ARTICLE{Ruzza2026,
       author = {{Ruzza}, Alessandro and {Lodato}, Giuseppe and {Rosotti}, Giovanni and {Armitage}, Philip and {Facchini}, Stefano and {Andrews}, Sean M. and {Bae}, Jaehan and {Barraza-Alfaro}, Marcelo and {Benisty}, Myriam and {Curone}, Pietro and {Fasano}, Daniele and {Hall}, Cassandra and {Hilder}, Thomas and {Izquierdo}, Andr{\'e}s F. and {Longarini}, Cristiano and {M{\'e}nard}, Fran{\c{c}}ois and {Pinte}, Christophe and {Stadler}, Jochen and {Teague}, Richard and {Terry}, Jason and {Wilner}, David J. and {Winter}, Andrew J. and {Yoshida}, Tomohiro C. and {Zawadzki}, Brianna},
        title = "{exoALMA. XXIII. Estimating Disk and Planet Properties from Dust Morphologies with DBNets 2.0}",
      journal = {\apjl},
         year = 2026,
        month = mar,
       volume = {1000},
       number = {1},
          eid = {L16},
        pages = {L16},
          doi = {10.3847/2041-8213/ae434c},
archivePrefix = {arXiv},
       eprint = {2603.13149},
 primaryClass = {astro-ph.EP},
       adsurl = {https://ui.adsabs.harvard.edu/abs/2026ApJ..1000L..16R}
}

@ARTICLE{Hardiman2026,
       author = {{Hardiman}, Caitlyn and {Pinte}, Christophe and {Price}, Daniel J. and {Hilder}, Thomas and {Hammond}, Iain and {Danilovich}, Ta{\"\i}ssa and {Andrews}, Sean M. and {Teague}, Richard and {Rosotti}, Giovanni and {Flock}, Mario and {Cataldi}, Gianni and {Bae}, Jaehan and {Barraza-Alfaro}, Marcelo and {Benisty}, Myriam and {Cuello}, Nicol{\'a}s and {Curone}, Pietro and {Czekala}, Ian and {Facchini}, Stefano and {Fasano}, Daniele and {Fukagawa}, Misato and {Galloway-Sprietsma}, Maria and {Garg}, Himanshi and {Hall}, Cassandra and {Huang}, Jane and {Ilee}, John D. and {Izquierdo}, Andres F. and {Kanagawa}, Kazuhiro and {Lesur}, Geoffroy and {Lodato}, Giuseppe and {Longarini}, Cristiano and {Loomis}, Ryan and {Menard}, Francois and {Orihara}, Ryuta and {Stadler}, Jochen and {Yen}, Hsi-Wei and {Fernandez}, Gaylor Wafflard- and {Wilner}, David J. and {Winter}, Andrew J. and {W{\"o}lfer}, Lisa and {Yoshida}, Tomohiro C. and {Zawadzki}, Brianna},
        title = "{exoALMA. XIX. Confirmation of Non-thermal Line Broadening in the DM Tau Protoplanetary Disk}",
      journal = {\apjl},
         year = 2026,
        month = feb,
       volume = {997},
       number = {2},
          eid = {L47},
        pages = {L47},
          doi = {10.3847/2041-8213/ae313a},
archivePrefix = {arXiv},
       eprint = {2602.01620},
 primaryClass = {astro-ph.EP},
       adsurl = {https://ui.adsabs.harvard.edu/abs/2026ApJ...997L..47H}
}

@ARTICLE{Flaherty2020,
       author = {{Flaherty}, Kevin and {Hughes}, A. Meredith and {Simon}, Jacob B. and {Qi}, Chunhua and {Bai}, Xue-Ning and {Bulatek}, Alyssa and {Andrews}, Sean M. and {Wilner}, David J. and {K{\'o}sp{\'a}l}, {\'A}gnes},
        title = "{Measuring Turbulent Motion in Planet-forming Disks with ALMA: A Detection around DM Tau and Nondetections around MWC 480 and V4046 Sgr}",
      journal = {\apj},
         year = 2020,
        month = jun,
       volume = {895},
       number = {2},
          eid = {109},
        pages = {109},
          doi = {10.3847/1538-4357/ab8cc5},
archivePrefix = {arXiv},
       eprint = {2004.12176},
 primaryClass = {astro-ph.SR},
       adsurl = {https://ui.adsabs.harvard.edu/abs/2020ApJ...895..109F}
}

@ARTICLE{Antilen2026,
       author = {{Antilen}, Juanita and {Pinilla}, Paola and {Li}, Dafa and {Villenave}, Marion and {Sierra}, Anibal and {Liu}, Yao and {Benisty}, Myriam and {Ginski}, Christian},
        title = "{Diverse dust vertical height and settling strength conditions in protoplanetary discs}",
      journal = {\mnras},
         year = 2026,
        month = jul,
       volume = {549},
       number = {3},
          eid = {stag882},
        pages = {stag882},
          doi = {10.1093/mnras/stag882},
archivePrefix = {arXiv},
       eprint = {2605.06904},
 primaryClass = {astro-ph.EP},
       adsurl = {https://ui.adsabs.harvard.edu/abs/2026MNRAS.549ag882A}
}

@ARTICLE{Woitke2016,
       author = {{Woitke}, P. and {Min}, M. and {Pinte}, C. and {Thi}, W.-F. and {Kamp}, I. and {Rab}, C. and {Anthonioz}, F. and {Antonellini}, S. and {Baldovin-Saavedra}, C. and {Carmona}, A. and {Dominik}, C. and {Dionatos}, O. and {Greaves}, J. and {G{\"u}del}, M. and {Ilee}, J.~D. and {Liebhart}, A. and {M{\'e}nard}, F. and {Rigon}, L. and {Waters}, L.~B.~F.~M. and {Aresu}, G. and {Meijerink}, R. and {Spaans}, M.},
        title = "{Consistent dust and gas models for protoplanetary disks. I. Disk shape, dust settling, opacities, and PAHs}",
      journal = {\aap},
         year = 2016,
        month = feb,
       volume = {586},
          eid = {A103},
        pages = {A103},
          doi = {10.1051/0004-6361/201526538},
archivePrefix = {arXiv},
       eprint = {1511.03431},
 primaryClass = {astro-ph.EP},
       adsurl = {https://ui.adsabs.harvard.edu/abs/2016A&A...586A.103W}
}

@software{Dominik2021,
       author = {{Dominik}, Carsten and {Min}, Michiel and {Tazaki}, Ryo},
        title = "{OpTool: Command-line driven tool for creating complex dust opacities}",
 howpublished = {Astrophysics Source Code Library, record ascl:2104.010},
         year = 2021,
        month = apr,
          eid = {ascl:2104.010},
archivePrefix = {ascl},
       eprint = {2104.010},
       adsurl = {https://ui.adsabs.harvard.edu/abs/2021ascl.soft04010D}
}

\appendix
\section{Simulation Setup Details}
\subsection{Temperature Structure}\label{app:TStruc}
\paragraph{Vertically Isothermal Temperature Structure}
The midplane temperature in our simulations is determined from the stellar parameters of a one-solar-mass pre-main sequence star based on the first recorded stellar structure of the \cite{Baraffe2015} evolutionary tracks.
\begin{equation}
    T(R,z)=T_\mathrm{mid}(R)= \left(\frac{ \varphi L_*}{4\pi  R^2 \sigma_\mathrm{SB}}\right)^{\nicefrac{1}{4}},
    \label{eq:Irrad}
\end{equation}
where $\varphi=0.05$ is the assumed angle at which the radiation is incident on the disk surface and $L_*$ is the stellar luminosity.

\paragraph{Temperature Structure from Flux-Limited Diffusion Models}
We have calculated disk temperatures via gray flux-limited radiative diffusion for our simulations using \texttt{pyFLD}\footnote{\url{https://github.com/alexziab/pyFLD/}; see \url{https://zenodo.org/records/22878208} for the exact version used in this work} with temperature stratification for the same stellar parameters and surface density profiles assumed for the vertically isothermal runs. The \texttt{pyFLD} model assumed a monodisperse $\mu$m-sized dust distribution with a uniform dust-to-gas ratio of $10^{-3}$, the opacities for which were computing using \texttt{optool} \citep{Dominik2021} with the DIANA standard composition \citep{Woitke2016}. Irradiation was included via frequency-dependent ray tracing from the central star.

The temperature structures have been fitted with an analytic function:
\begin{align}
    h^2 &= {h_\mathrm{surf}^2 + (h_\mathrm{surf}^2 - h_\mathrm{mid}^2)\left[\exp\left(-\exp\left\{-\frac{\left|\frac{z}{h_\mathrm{mid}}\right| - z_\mathrm{trans} \left(\frac{R}{\si{\AU}}\right)^{-0.09}}{w_\mathrm{trans} \left(\frac{R}{\si{\AU}}\right)^{-0.05}}\right\}\right) - 1\right]} \\
    T &= \frac{\mu m_\mathrm{p} h^2}{k_\mathrm{B}} v_\mathrm{K}^2,
\end{align}
where $z_\mathrm{trans}=4.6\, H$ is the height at which temperatures switch from cold midplane to hot atmosphere and $w_\mathrm{trans}=0.27$ is the width of the transition zone. 
Furthermore, $h_\mathrm{mid}$ determines the midplane temperature (from \autoref{eq:Irrad}) and $h_\mathrm{surf}$ is a fit to the surface temperature, given by
\begin{equation}
    h_\mathrm{surf} = \sqrt{2.4} h_0 \left(\frac{R}{\si{\AU}}\right)^{0.28},
\end{equation}
where $h_0$ is the midplane aspect ratio ($h_\mathrm{mid}$) at \SI{1}{\AU}.
This structure is then used as an input for the hydrostatic integrator (see next section) and as an equilibrium profile for the subsequent hydrodynamic simulations.

We note that different methods for radiative transfer lead to different temperature profiles (see \autoref{app:TstrucRADMC}). This is expected, as even though both codes utilize frequency-dependent stellar irradiation, \texttt{pyFLD} uses a gray (i.e., frequency-integrated) approach for radiative transfer. We use this specific profile only to demonstrate the potential effects of temperature stratification in general. 

\subsection{Density Structure}
\label{app:DenStruc}
\paragraph{Vertically Isothermal Simulations}
In the vertically isothermal simulations, we define the initial condition as the standard hydrostatic circumstellar disk, with a \cite*{LyndenBell1974} midplane density profile
\begin{align}
    \rho_\mathrm{g,mid} &=\frac{M_{\text{d}}(1+\beta_{\Sigma})}{(2\pi)^{\nicefrac{3}{2}} H_\mathrm{g} R_\mathrm{c}^2} \left(\frac{R}{R_\mathrm{c}}\right)^{\beta_{\Sigma}}\exp{\left[-\left(\frac{R}{R_\mathrm{c}}\right)^{2+\beta_{\Sigma}}\right]} \label{eq:rhomid} \\ 
    \rho_\mathrm{g} &= \rho_\mathrm{g,mid} \exp{\left[\left(\frac{H_\mathrm{g}}{R}\right)^{-2}\left(\frac{R}{\sqrt{R^2+z^2}} - 1\right)\right]}, \label{eq:rhoz}
\end{align}
where $R_\mathrm{c}$ is the disk's characteristic radius, and $\beta_\Sigma=-1$ is the column density power-law exponent.
The disk's rotation profile then follows from hydrostatic equilibrium as
\begin{align}
\frac{\Omega^2(R,z)}{\Omega_\mathrm{K}^2} = &\left(\frac{H_\mathrm{g}}{R}\right)^2\left(\beta_T+\beta_\rho-(\beta_\Sigma + 2)\left(\frac{R}{R_\mathrm{c}}\right)^{\beta_\Sigma+2}\right) \nonumber \\ 
&- \frac{\beta_T R}{\sqrt{R^2+z^2}} + \beta_T + 1 , \label{eq:rot}
\end{align}
where $\beta_T=-0.5$ is the temperature power-law exponent.
\paragraph{Simulations with Vertical Temperature Structure}
No analytical solution exists for the hydrostatic structure of a disk with vertical temperature stratification, as described above. 
To derive an equilibrium structure for the initial condition in these simulations, we instead numerically integrate the hydrostatic equations in spherical coordinates, following the same procedure as \cite{Flock2013} and \cite{Fuksman2022}.
For this, integration begins in the disk midplane, given the initial condition from \autoref{eq:rhomid} and \autoref{eq:rot} (for $z=0$).
We are then integrating the following ordinary differential equation one step away from the midplane on the desired spatial grid ($\mathrm{d}\vartheta$) using the midpoint integration rule
\begin{equation}
    \frac{\mathrm{d}\report{\ln}(\rho_\mathrm{g})}{\mathrm{d}\vartheta} = \frac{1}{\tan(\vartheta)}\frac{v_\mathrm{\varphi}^2}{c_\mathrm{s}^2} - \frac{1}{c_\mathrm{s}^2}\frac{\mathrm{d}c_\mathrm{s}^2}{\mathrm{d}\vartheta}.
\end{equation}
Given the new density structure at $\vartheta_0+\mathrm{d}\vartheta$, we then compute the azimuthal velocity component in the respective cells.
\begin{equation}
    \frac{v_\varphi^2}{r} = \frac{\mathrm{d\Phi}}{\mathrm{d}r} + \frac{\mathrm{d}c_\mathrm{s}^2}{\mathrm{d}r} + c_\mathrm{s}^2 \frac{\mathrm{d}\report{\ln}(\rho_\mathrm{g})}{\mathrm{d}r},
\end{equation}
where $r$ is the radius in spherical coordinates.
The resulting density and rotation profiles are in force equilibrium.

\section{Dust Opacity Tables for \athena{} with \tripod{}}\label{app:OpTab}
\begin{figure}
    \centering
    \includegraphics[width=0.5\linewidth]{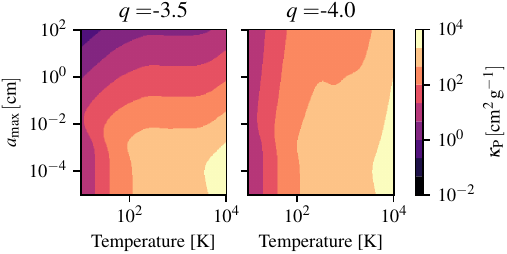}
    \caption{Slices through our three-dimensional dust opacity tables for the Planck opacities (function of $a_\mathrm{max}$, $q$, and $T$), used in all hydrodynamic simulations.}
    \label{fig:OpaTab}
\end{figure}

We calculate three-dimensional dust opacity tables for the Planck opacities (as a function of $a_\mathrm{max}$, $q$, and $T$), used for all hydrodynamic simulations. We use the \texttt{dsharp\_opac} Python module with the default DSHARP dust properties and optical constants. We show four slices through these tables in \autoref{fig:OpaTab}.
At every simulation timestep and in every grid cell, we perform a trilinear interpolation on these tables to find the opacity for the given maximum grain size, power-law exponent, and temperature in that grid cell. The algorithm used is identical to the interpolation method provided in the \texttt{growpacity} python package \citep{growpacity}. 

\section{\texttt{RADMC-3D} Model Structure and Opacities}
\label{app:TstrucRADMC}
\begin{figure*}[ht]
\centering
    \includegraphics[height=0.4\textwidth]{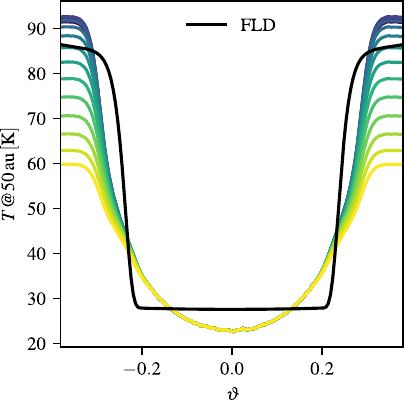}
    \includegraphics[height=0.4\textwidth]{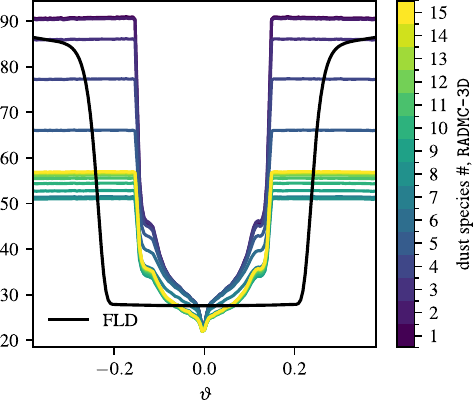}
    \caption{Temperature profiles at \SI{50}{\AU} distance to star calculated with \radmc{} (colored lines for different dust species), and temperature profile assumed in the hydrodynamic simulation. Species 1 corresponds to the smallest grains, species 15 to the largest. The left panel corresponds to the initial condition, the right panel corresponds to the last simulation snapshot (after 1500 orbits)}
    \label{fig:RADMC_Temp}
\end{figure*}

\begin{figure}[ht]
\centering
    \includegraphics[width=0.5\textwidth]{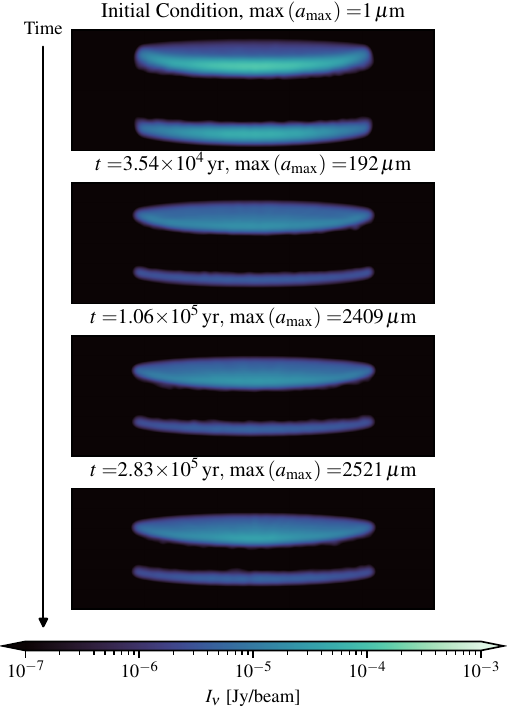}
    \caption{Equivalent to \autoref{fig:RADMC_Timeseries} but for scattered light at \SI{1.6}{\micro \meter}.}
    \label{fig:scatterseries}
\end{figure}

Here we show the temperature structures and optical depth of the \radmc{} models presented in \autoref{fig:RADMC_Timeseries}. The left panel of \autoref{fig:RADMC_Temp} shows the temperature structure at \SI{50}{\AU}, calculated for the initial condition---that is a MRN size distribution between \SI{0.1}{\mu m} and \SI{1}{\mu m} with a constant total dust-to-gas ratio of \SI{1}{\percent} in every grid cell. For comparison, we plot the temperature profile, used throughout the hydrodynamic simulation with vertical stratification, in black. This profile is based on a flux-limited diffusion model. It can be clearly seen that, for once, the vertical structures differ significantly in terms of the overall profile, but also across the dust species in the \radmc{} simulation. This illustrates the necessity for a more self-consistent future study of VSI with radiative transfer.
Furthermore, the \radmc{} temperature structure changes significantly over the simulation's runtime, as shown in the right panel, which depicts the profile after \SI{1500}{orbits}, calculated from the respective dust size distributions of the hydrodynamic simulation.
The dust has significantly sedimented toward the midplane, meaning that the scattering surface has also moved toward the midplane (see \autoref{fig:scatterseries}). The highly concentrated dust layer in the midplane reaches high optical depth which leads to an additional decrease of the temperature at this location.

We therefore note that our simulation with temperature structure only represent one possible configuration, and that more self-consistent approaches will be necessary in the future.

\begin{figure*}[ht]
    \includegraphics[width=\textwidth]{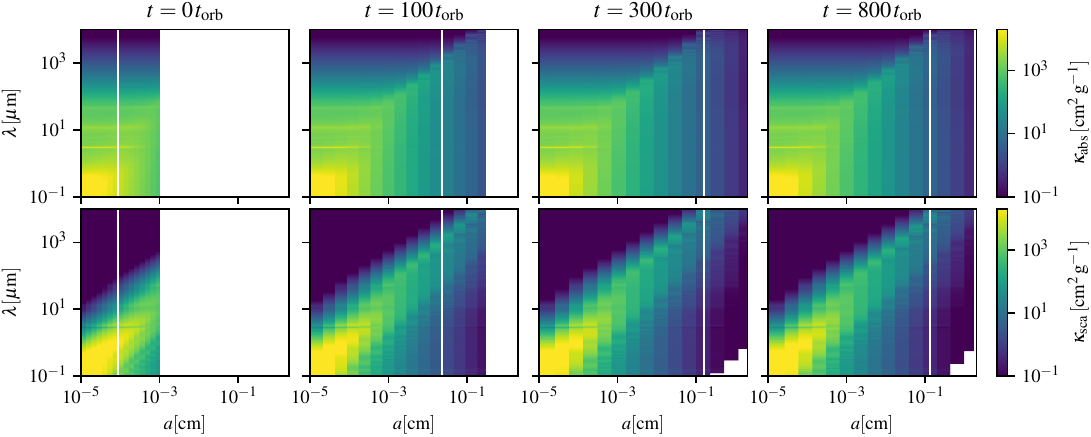}
    \caption{Opacity tables for the four \radmc{} simulations depicted in \autoref{fig:RADMC_Timeseries}. White vertical lines mark the max.\ particle size.}
    \label{fig:Opacities}
\end{figure*}

\begin{figure}
    \centering
    \includegraphics[width=0.5\textwidth]{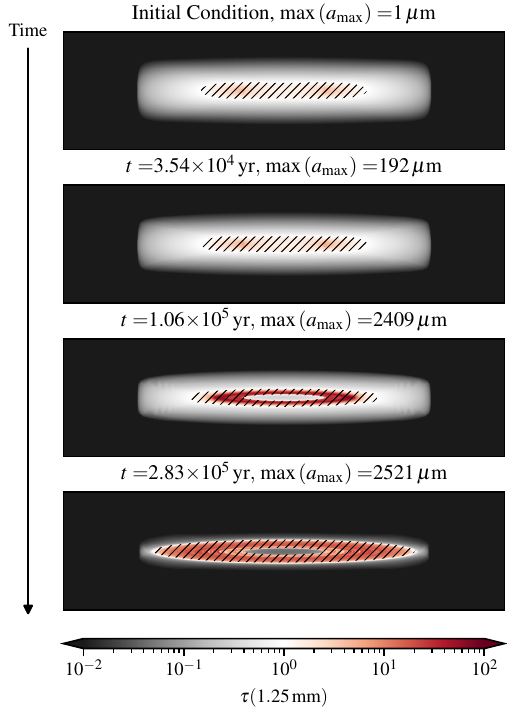}
    \caption{Optical depth of the observations in \autoref{fig:RADMC_Timeseries}. Hatched areas mark regions with $\tau>1$.}
    \label{fig:opticaldepth}
\end{figure}

We also present the respective opacity tables in \autoref{fig:Opacities}. Since the dust grows over time, also the opacities used for the \radmc{} simulation are different for the four snapshots in \autoref{fig:RADMC_Timeseries}. Furthermore, \autoref{fig:opticaldepth} depicts the respective optical depth at \SI{1.25}{\milli \meter} corresponding to \autoref{fig:RADMC_Timeseries}.

\section{Additional Simulations} 
\subsection{Higher Fragmentation Velocity}
\label{app:vfrag}
\begin{figure*}[ht]
\centering
    \includegraphics[width=\textwidth]{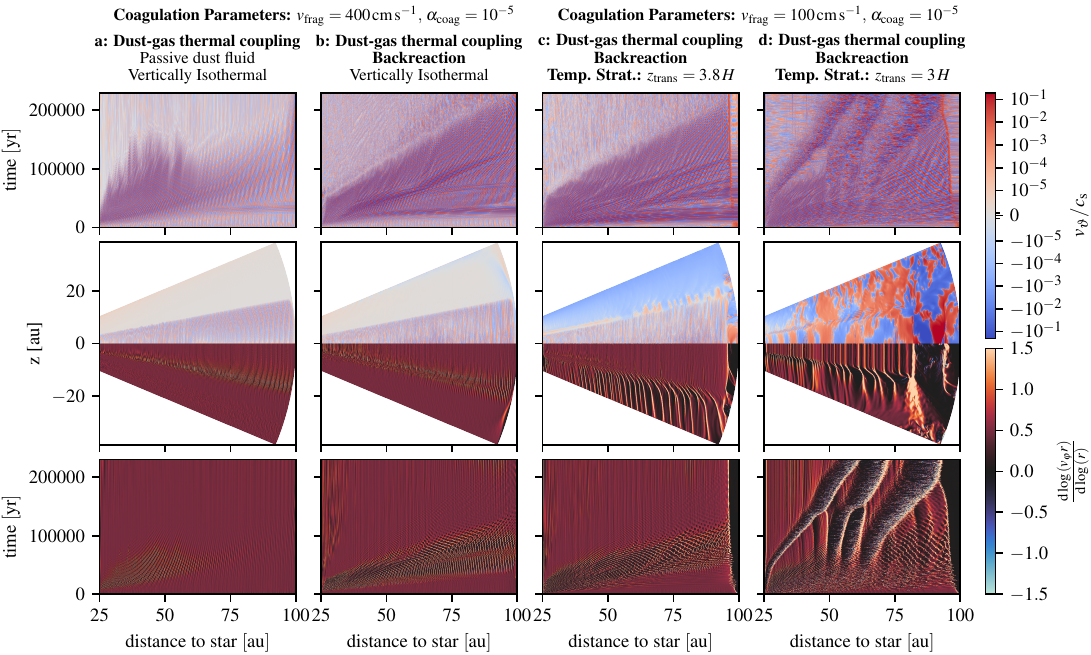}    
    \caption{Same as \autoref{fig:vfr100_gas} but for the four additional simulations. Columns (a) and (b) are the same simulation setup as columns (b) and (c) in \autoref{fig:vfr100_gas} and but with $v_\mathrm{frag}=\SI{400}{\centi \meter \per \second}$. Columns (c) and (d) show simulations with different temperature stratification: in column (c), the transition height has been reduced to $3.8\, H$, in column (d) it has been lowered to $3\, H$.}
    \label{fig:ExtraSims_gas}
\end{figure*}

\begin{figure*}[ht]
\centering
    \includegraphics[width=\textwidth]{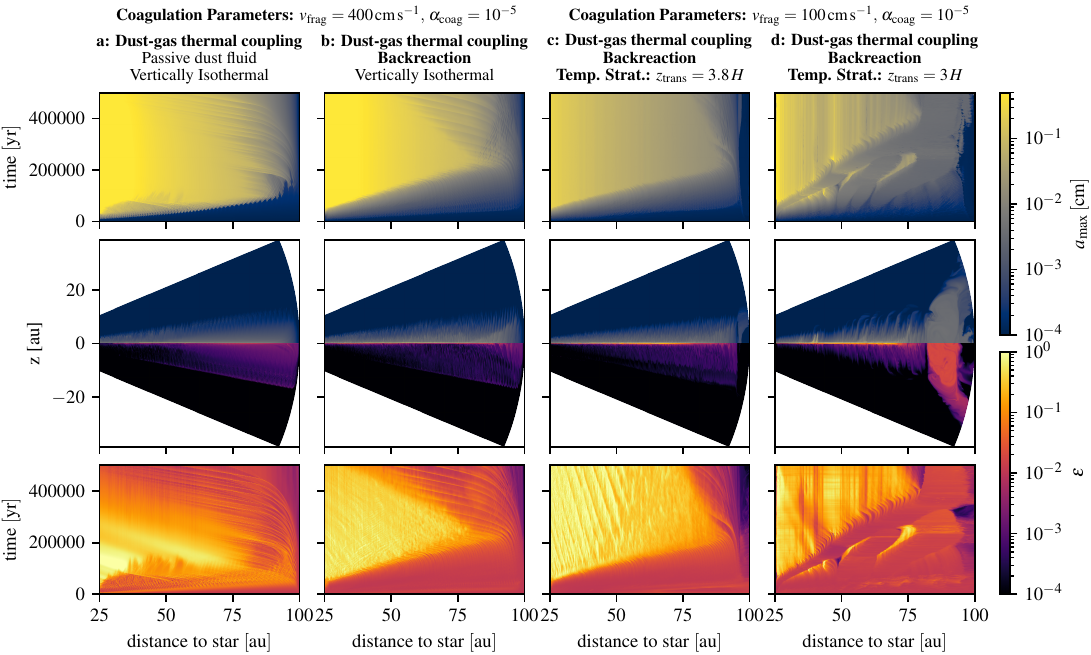}    
    \caption{Same as \autoref{fig:vfr100_dust} but for the four additional simulations. Columns (a) and (b) are the same simulation setup as columns (b) and (c) in \autoref{fig:vfr100_dust} and but with $v_\mathrm{frag}=\SI{400}{\centi \meter \per \second}$. Columns (c) and (d) show simulations with different temperature stratification: in column (c), the transition height has been reduced to $3.8\, H$, in column (d) it has been lowered to $3\, H$.}
    \label{fig:ExtraSims_dust}
\end{figure*}

Here, we present two simulations with a four times higher fragmentation velocity than the default value of \SI{100}{\centi\meter\per\second}, that was used throughout this letter.
The result is shown in \autoref{fig:ExtraSims_gas} and \autoref{fig:ExtraSims_dust}.
Column (a) shows a simulation with thermal relaxation but without backreaction. 
While this setup results in a steadily turbulent disk for $v_\mathrm{frag}=\SI{100}{\centi\meter\per\second}$, we find that for the $v_\mathrm{frag}=\SI{400}{\centi\meter\per\second}$ case, turbulence gets fully suppressed even if backreaction is neglected. This is consistent with previous findings by \cite{Fukuhara2024} and the simulations by \cite{Pfeil2024}, that show that, if cooling times are too long, VSI cannot sustain the dust layer against sedimentation. 

The result in column (b) is equivalent to the vertically isothermal simulation with thermal relaxation and dust backreaction (i.e., column (c) in \autoref{fig:vfr100_gas} and \autoref{fig:vfr100_dust}), except for the higher fragmentation velocity. As can be seen, the result is qualitatively similar but with significantly larger particles. We can see the same highly sedimented midplane layer and no significant VSI turbulence.  

These simulations show that the results presented in the main part of this letter represent a favorable case for the VSI and affirm our finding that coagulation, sedimentation, slow cooling, and backreaction can terminate the VSI activity within a fraction of the disk lifetime. 

\subsection{Temperature Transitions Closer to the Midplane}
\label{app:TstrucSims}
As discussed in \autoref{sec:dissT}, moving the temperature transition to lower parts of the disk atmosphere can significantly enhance the VSI activity \citep[see also][]{Yun2025,Yun2025a}. 
The results are shown in columns (c) and (d) of \autoref{fig:ExtraSims_gas} and \autoref{fig:ExtraSims_dust}. If the temperature transition is moved only slightly down, as shown in panel c, no significant changes occur. The dust is sedimenting past the transition before the VSI can cause significant turbulence. 

If the transition height is lowered to $3 \, H$, where dust remains present for longer amounts of time, VSI can cause significant turbulence in the atmosphere, which also strongly perturbs the midplane dust layer. 
Nonetheless, dust growth and sedimentation proceed toward the midplane. 
Note that in this simulation, due to the initially very strong VSI turbulence, large bands of constant angular momentum form (see \autoref{fig:ExtraSims_gas}, column d) that grow into large vortices. This behavior typically occurs in axisymmetric simulations and is not to be expected in three-dimensional simulations. 
Despite this very strong turbulence, the disk becomes mostly quiescent after \SI{400}{\kilo \years}.

\end{CJK*}
\end{document}